\documentclass{iopart}       % arXiv version
\usepackage{iopams}      
\usepackage{setstack}
\usepackage[all]{xy}

\def\IBSsecondorder{I.1}

\def\IHepsilondef{I.11}
\def\IHepsilonode{I.13}
\def\IGepsilondef{I.14}
\def\IHsubepsilonode{I.16}
\def\IHsubnepsilondef{I.17}
\def\IZsubepsilonode{I.19}
\def\IDDepsilondef{I.24}
\def\IaltDDepsilondef{I.25}
\def\IHsubzeroepssoln{I.28}
\def\IKzerosoln{I.29}
\def\IKtwosoln{I.32}
\def\Ifzerodef{I.35}
\def\IZdef{I.37}
\def\IAphiepsilondef{I.43}
\def\IHzeroepsaa{I.44}
\def\IUhatepsilonode{I.45}
\def\IUndef{I.47}
\def\IMneqns{I.51}
\def\IMzerosoln{I.52}
\def\IMtwoKtwoavg{I.54}
\def\IGtwoeqn{I.55}
\def\IMtworearrangeavg{I.56}
\def\IVhatdef{I.65}
\def\Icontractionsign{I.A.3}
\def\IMoyalcommutator{I.A.6}
\def\Icanonicaltransformations{I.3.3}
\def\Inotationepsilonderiv{I.3.4}
\def\Iactionoperators{I.5.2}
\def\IMoyalnotation{Appendix~I.A}
\def\Ifunctionnotation{Appendix~I.B}
\def\IZissmooth{Appendix~I.C}
\def\Inotationaveraging{Appendix~I.D}
\def\IGexists{Appendix~I.E}
\def\Iantideriv{Appendix~I.G}
\def\IuniquenessofK2{Appendix~I.H}

\def\metaplecticcovariance{6.27}

\def\amatrix{a}

\begin{document}
\title{Multidimensional quantum normal forms, Moyal star product, and
torus quantization}
\author{Matthew Cargo, Alfonso Gracia-Saz and R G Littlejohn}
\address{Departments of Physics and Mathematics, University of
California, Berkeley, California 94720 USA}

\ead{robert@wigner.berkeley.edu}

\begin{abstract}

A normal form transformation is carried out on the operators of a
complete set of commuting observables in a multidimensional,
integrable quantum system, mapping them by unitary conjugation into
functions of the harmonic oscillators in the various degrees of
freedom.  The transformation works at the level of the Weyl symbols of
the operators, which are manipulated as formal power series in $\hbar$
by use of the Moyal star product.  It is assumed that the Weyl symbol
of one of the operators (the Hamiltonian) has a generic, stable fixed
point in phase space.  The normal form transformation takes place in a
neighborhood of this fixed point.  Once the normal form has been
achieved, the Einstein-Brillouin-Keller or torus quantization rule
follows easily, including higher order corrections in $\hbar$.
Crucial parts of the normal form transformation are not obvious
generalizations of the one-dimensional case, nor is final quantization
rule.  The result raises some issues of differential geometry not
found in the one-dimensional case.

\end{abstract}

\pacs{03.65.Sq, 02.40.Gh, 03.65.Vf, 02.20.Sv}

%\maketitle %use this if you want to have title on separate page

\section{Introduction}

In a previous article (Cargo \etal 2004), hereinafter referred to as
I, we have developed a normal form transformation for a
one-dimensional quantum Hamiltonian whose Weyl symbol has a generic,
stable fixed point in phase space.  In particular, we showed that such
a Hamiltonian can be transformed via unitary conjugation into a
function of the harmonic oscillator in a neighborhood of the fixed
point, in the sense of formal power series in $\hbar$.  One of the
consequences is the Bohr-Sommerfeld rule for the eigenvalues of the
Hamiltonian, extended to higher order in $\hbar$.  In this article we
generalize this normal form transformation to multidimensional,
integrable systems, in which a complete set of commuting observables,
one of which has a generic, stable fixed point in phase space, is
transformed via unitary conjugation into a set of functions of the
harmonic oscillators in the various degrees of freedom.  One
consequence is the Einstein-Brillouin-Keller (EBK) or torus
quantization rule, including higher order corrections in $\hbar$.
Part of this calculation is a straightforward generalization of the
methods of I, and part involves nontrivial new techniques.  In the
following a familiarity with I is assumed.  References to equations,
sections and appendices of I are preceded by ``I'' (for example,
Equation~(\IHsubnepsilondef), Section~\Inotationepsilonderiv, or
\Ifunctionnotation).

In our opinion the normal form transformation itself, which has many
variations not yet explored (unstable fixed points, regions including
separatrices, nongeneric fixed points, etc.), and the diagrammatic and
other techniques developed for carrying out the normal form
transformation, are the most important results of I and of this paper.
Nevertheless, the torus quantization rule and its higher order
corrections are the results that are most easily appreciated and
understood, so we have given them particular prominence in the
following discussion.

The usual torus quantization rule (Einstein, 1917; Brillouin, 1926;
Keller, 1958; Percival, 1973; Berry and Tabor, 1976) is valid only for
integrable systems, which at the level of the usual formula means that
the classical Hamiltonian is integrable in a classical sense.  If $H$
is a classical, integrable Hamiltonian of $N$ degrees of freedom with
actions $A=(A^1, \ldots, A^N)$ and angles $\phi=(\phi^1, \ldots,
\phi^N)$, then the torus quantization rule states that the eigenvalues
of the corresponding quantum Hamiltonian $\hat H$ are given
approximately by $E_n = H(A)$, where $H$ is expressed as a function of
the actions, where the actions are evaluated at $A^j = (n_j +
\gamma_j/4)\hbar$, where $n=(n_1, \ldots, n_N)$ is a vector of integer
quantum numbers, and where $\gamma=(\gamma_1, \ldots, \gamma_N)$ is
the vector of Maslov indices (Maslov and Fedoriuk 1981, Littlejohn and
Robbins 1987, Eckhardt and Wintgen 1991, de Gosson 1997, Foxman and
Robbins 2004).  See \ref{notation} for notation.  The torus
quantization rule is discussed more fully and precisely in the
explanation of (\ref{theresult}) below.

Examples of integrable systems include central force motion in any
number of dimensions in ordinary particle mechanics, objects such as
the $3nj$-symbols in angular momentum theory (which are the
eigenfunctions of complete sets of commuting observables), and
integrable lattice and soliton systems.  Integrable systems have also
exerted a powerful influence on various areas of mathematics and
mathematical physics, including symplectic geometry and representation
theory (Weinstein 1977, Marsden and Ratiu 1999, Kirillov 1976).  They
are also the reference point for understanding partially integrable or
nonintegrable systems.

A normal form transformation of the kind developed in I has been
studied by Helffer and Robert (1984) in the one-dimensional case.
These authors use Fourier integral operators to carry out the normal
form (rather than the Moyal star product, as in our work), and they
prove the existence of the asymptotic $\hbar$ series for the
eigenvalues.  They do not, however, explicitly compute the higher
order terms.  In related work, Charbonnel (1986) has studied a
multidimensional example.  

One question that arises immediately when contemplating higher order
corrections to the torus quantization rule is what shall be the
definition of an integrable system in quantum mechanics.  A set of $N$
operators need only commute through order $\hbar$ for their principal
symbols to Poisson commute, and thereby to define a classically
integrable system.  The latter is all that is required for the usual
(lowest order) torus quantization formula to be meaningful.  When we
extend torus quantization to higher order in $\hbar$, is Poisson
commutativity of the principal symbols sufficient, or must the
operators commute to higher order in $\hbar$?  One might suppose that
terms that are higher order in $\hbar$ could be handled by some kind
of perturbation theory, and that classical integrability would
suffice.

In fact, it turns out that for torus quantization to be meaningful to
all orders in $\hbar$ it is necessary to have a complete set of
operators that commute to all orders in $\hbar$.  Without this
condition, the normal form transformation that is the basis of our
analysis does not exist, at least in the form of the power series in
$\hbar$ that we posit for it.  If this and other, reasonable (generic)
conditions to be described below hold, then as we shall show the torus
quantization rule can be extended to arbitrary order in $\hbar$.

We now present the generalized torus quantization rule, deferring the
details of our assumptions to the main body of the paper.  Let $({\hat
H}^1, \ldots, {\hat H}^N)$ be a set of $N$ commuting, Hermitian
operators, $[{\hat H}^j, {\hat H}^k]=0$, $j,k = 1,\ldots, N$, of which
${\hat H}^1$ is ``the'' Hamiltonian (above referred to as simply
${\hat H}$).  Let the (Weyl) symbols of these operators be $(H^1,
\ldots, H^N)$ (that is, without the hat.  See \ref{notation}.)  These 
are assumed to have an $\hbar$-expansion of the form,
	\begin{equation}
	H^j = H^j_0 + \hbar^2 H^j_2 + \hbar^4 H^j_4 + \ldots,
	\label{Hjhbarexpansion} 
	\end{equation} 
that is, they are ``slowly varying'' or ``admissible'' (Voros 1977)
with only even powers of $\hbar$.  The leading terms $H^j_0$ are the
``principal symbols'' of the operators ${\hat H}^j$.  We assume that
odd powers of $\hbar$ are absent from this expansion because their
inclusion serves only to complicate all the formulas of this analysis
without introducing any fundamentally new issues.  In addition, most
of the examples we can think of (such as $({\hat H}, {\hat L}^2, {\hat
L}_z)$ in a 3-dimensional central force problem) have symbols that are
even series in $\hbar$, a property that evidently is related to the
Hermiticity of the operators ${\hat H}^j$.  The principal symbols
$H^j_0$ Poisson commute, $\{H_0^j, H_0^k\}=0$, since all correction
terms in the Moyal series (\IMoyalcommutator) for the symbol of a
commutator are of higher order in $\hbar$. 

The principal symbol $H^1_0$ of the quantum Hamiltonian ${\hat H}^1$
will be treated as the classical Hamiltonian (above referred to
simply as $H$).  We assume that it possesses a generic extremum at
some (fixed) point of phase space, so that classical motion in a
neighborhood of this point is bounded and the fixed point is stable.
Action-angle variables $(\phi,A)$ are defined in this neighborhood,
and the principal symbols can be expressed as functions of the
actions, $H^j_0 = H^j_0(A)$.

Then the eigenvalues $E^j_n$ of ${\hat H}^j$ are given by
	\begin{eqnarray}
	E^j_n = & \Biggl.\Biggl[ H^j_0(A) + \hbar^2 \Biggl<
	H^j_2 -\frac{1}{48} \omega_{jk} 
	\xymatrix@C=1pc{(D^\mu \ar[r] & 
	A^k \ar@<0.5ex>[r] \ar@<-0.5ex>[r]
	& D_\mu)} 
	+\frac{1}{16} \omega_{jkl}
	\xymatrix@C=1pc{(A^k \ar@<0.5ex>[r] 
	\ar@<-0.5ex>[r] & A^l)}
	\nonumber \\
	& -\frac{1}{24} \omega_{jklm} 
	\xymatrix@C=1pc{(A^k \ar[r] & A^l \ar[r] & A^m)}
	\Biggr>_\phi 
	+ O(\hbar^4) \Biggr] 
	\Biggr|_{A=(n+\gamma/4)\hbar},
	\label{theresult}
	\end{eqnarray}
where $n$ and $\gamma$ are $N$-vectors of integer quantum numbers and
Maslov indices, as above.  The ranges of the integers $n_j$ depend on
the contours on the tori used to defined the actions $A^j$, in a
manner that will be made precise in subsection~\ref{secondstage}.
Equation~(\ref{theresult}) is the torus quantization rule, carried
through order $\hbar^2$.

Equation~(\ref{theresult}) uses the following notation.  (See also
\ref{notation}.)  Repeated indices are summed.  The frequency
matrix of the principal symbols and its action derivatives are denoted by
	\begin{equation}
	\omega_{jk} = \frac{\partial H^j_0}
	{\partial A^k},
	\qquad
	\omega_{jkl} = \frac {\partial \omega_{jk}}{\partial A^l},
	\qquad
	\omega_{jklm} = \frac{\partial^2 \omega_{jk}}{\partial A^l
	\partial A^m}. 	
	\label{omegajkdef}
	\end{equation}
These are all functions of the actions $A$.  The angle bracket
$\langle\;\rangle_\phi$ at order $\hbar^2$ in (\ref{theresult})
indicates an average over the angles $\phi$ (a torus average).  The
quantities inside the angle bracket depend on both $A$ and $\phi$, but
of course only on $A$ after the averaging.  The diagrammatic notation
is explained in \IMoyalnotation, with obvious generalizations to the
multidimensional case.  The notation $D^\mu$ stands for the
action-angle variables collectively, specifically,
	\begin{equation}
	\eqalign{
	D^\mu &= (\phi^1, \ldots, \phi^N, A^1, \ldots, A^N) \\
	D_\mu &= J_{\mu\nu}\, D^\nu = 
	(-A^1, \ldots, -A^N, \phi^1, \ldots, \phi^N)}
	\label{Dmudef}
	\end{equation}
(see \ref{notation}).  Finally, the entire expression is evaluated at
the quantized values of the actions, $A^j=(n_j + \gamma_j)\hbar$.

Equation~(\ref{theresult}) may be compared to its one-dimensional
counterpart (\IBSsecondorder).  The one-dimensional quantization rule at
order $\hbar^2$ is expressed in terms of a single diagram, the second
Moyal bracket $\{H, H\}_2 = \xymatrix@C=1pc{H \ar@<0.5ex>[r]
\ar@<-0.5ex>[r] & H}$ of $H$ with itself.  If $H$ is expressed as a
function of $A$ (the action of the one-dimensional problem), this one
diagram can be expanded by the chain rule yielding several diagrams,
as in the multidimensional formula (\ref{theresult}).
Equation~(\ref{theresult}), on the other hand, cannot in general be
condensed into diagrams involving the symbols $H^j$ alone, but rather
must be expressed in terms of diagrams involving the actions $A^j$,
and, in the case of the first diagram, also the angles $\phi^j$.  It
does reduce to (\IBSsecondorder) in the one-dimensional case, but it
is not a simple or obvious generalization of the one-dimensional
formula.  Equation~(\ref{theresult}) will be discussed more fully in
Sec.~\ref{discussion}, but for now we simply remark that of the three
diagrams shown, the final two are in a sense ``trivial'' consequences
of the formula for the symbol of a function of an operator, while the
first one is the difficult one that requires new techniques for its
evaluation.

The normal form transformation of this paper takes place on three
levels, of which the first two are purely classical.  Normal forms of
classical Hamiltonian systems have received considerable attention in
recent years (for example, Eliasson 1990, Fomenko 1991).  The first
level takes place in a small neighborhood of the fixed point of the
classical Hamiltonian $H^1_0$, in which the functions $H^j_0$ are
represented by their quadratic approximations about the fixed point
and a linear canonical transformation is used to achieve a classical
normal form.  (Specifically, it transforms these quadratic
approximations into linear combinations of harmonic oscillators.)  At
the next level, this linear canonical transformation in a small
neighborhood of the fixed point is extended to a nonlinear canonical
transformation in a certain ``region of interest,'' in which the level
sets of the functions $H^j_0$ are tori, achieving a nonlinear
classical normal form in that region.  (Specifically, the $H^j_0$ are
transformed into nonlinear functions of the harmonic oscillators).  At
the third level, the nonlinear classical normal form transformation is
used as the skeleton for the construction of a quantum normal form
transformation, valid in the sense of formal power series in $\hbar$,
constructed out of the Moyal star product.  (Specifically, the symbols
of the operators ${\hat H}^j$ become functions of the symbols of the
quantum harmonic oscillators in the region of interest).  The quantum
normal form then allows an easy derivation of the torus quantization
rule.

This paper is organized as follows.  Section~\ref{thecalculation}
contains an overview of the calculation, allowing readers who are
familiar with I to understand the main ideas and how they differ from
I.  Section~\ref{details} contains a more detailed discussion of
various steps in the presentation, including the rich subject of the
classical normal form.  Section~\ref{discussion} contains a discussion
of various aspects of the result (\ref{theresult}), especially certain
issues of differential geometry raised by the result.
Section~\ref{conclusions} contains conclusions.  Finally, a series of
appendices contain much of the real work of the calculation, as well
as a detailed presentation of an example.  

\section{An overview of the calculation}
\label{thecalculation}

In this section we present an overview of the calculation, assuming a
familiarity with I, and we point out various similarities and
differences with I.

One difference concerns the notation for Weyl symbols and their
expansions in powers of $\hbar$.  In I, the classical Hamiltonian was
taken to be the symbol $H$ of the quantum Hamiltonian $\hat H$, that
is, the whole symbol, possibly with an $\hbar$ dependence, whereas in
this paper the classical Hamiltonian is taken to be the principal
symbol $H^1_0$ of the quantum Hamiltonian ${\hat H}^1$, and similarly
we use the principal symbols $H^j_0$ of the ``other Hamiltonians''
together with $H^1_0$ to define a classically integrable system.
These principal symbols are independent of $\hbar$.  This approach is
necessary in the multidimensional case because the full symbols $H^j$
of the operators ${\hat H}^j$ do not Poisson commute in general,
except when $N=1$.  A consequence is that in this paper notations for
functions on phase space such as $H^j_n$, $K^j_n$, $G_n$, etc., are
independent of $\hbar$, whereas in I the analogous notations were
allowed to have an $\hbar$ dependence.

As in I, we begin by performing a canonical scaling of the operators
${\hat x}_j$ and ${\hat p}_j$ to give them both units of $\hbox{\rm
action}^{1/2}$.  This allows us to use a symmetrized form in
(\ref{Ijdef}) below for the harmonic oscillator actions, and it means
that all subsequent transformations preserve physical units.  See
\ref{normalformexample} for an example.

\subsection{The two stages}

The main part of the calculation is to transform the set of operators
$\{ {\hat H}^j \}$ via unitary conjugation into a new set of operators
$\{ {\hat M}^j \}$ that are functions of the harmonic oscillators $\{
{\hat I}^j \}$ in the various degrees of freedom.  The harmonic
oscillators and their symbols are given by
	\begin{equation}
	{\hat I}^j = \frac{1}{2}({\hat x}_j^2 + {\hat p}_j^2),
	\qquad
	I^j(z) = \frac{1}{2}(x_j^2 + p_j^2)
	\label{Ijdef}
	\end{equation}
(see \ref{notation}).  The ${\hat I}^j$ or $I^j$ are standard harmonic
oscillators of unit frequency, which are their own actions.  The new
operators ${\hat M}^j$ are considered functions of the harmonic
oscillators ${\hat I}^j$ if their symbols $M^j$ are functions of the
symbols $I^j$ in a neighborhood of the generic, stable fixed point of
the classical Hamiltonian $H^1_0$, in the sense of formal power series
in $\hbar$.  We denote the assumed fixed point by $z_0$.  Here we rely
on the theorem, the one-dimensional analog of which was discussed in
I, that an operator is a function of the operators ${\hat I}^j$ if and
only if its symbol is a function of the symbols $I^j$ (see
\ref{hosymbols}).

As in I, the transformation from the operators ${\hat H}^j$ to
operators ${\hat M}^j$ proceeds in two stages.  In the first stage
(the ``preparatory transformation'') we seek a unitary transformation
${\hat U}$ such that the operators ${\hat K}^j$, defined by
	\begin{equation}
	{\hat K}^j = {\hat U} {\hat H}^j {\hat U}^\dagger,
	\label{Kjdef}
	\end{equation}
have principal symbols $K^j_0$ that are functions of the harmonic
oscillator actions $I^j$.  That is, the symbol $K^j$ of the operator
${\hat K}^j$ is required to be a function of the harmonic oscillator
actions $\{I^j\}$ to lowest order in $\hbar$.  

In the second stage, we seek a sequence of near-identity unitary
operators ${\hat U}_2$, ${\hat U}_4$, ${\hat U}_6$, etc., such that
the operators ${\hat M}^j$, defined by
	\begin{equation}
	{\hat M}^j = {\hat V} {\hat H}^j {\hat V}^\dagger,
	\label{Mjdef}
	\end{equation}
where ${\hat V} = \ldots {\hat U}_6 {\hat U}_4 {\hat U}_2 {\hat U}$ as
in (\IVhatdef), have symbols $M^j$ that are formally functions of the
harmonic oscillator actions $I^j$ to all orders in $\hbar$.  Each
${\hat U}_n$ is given in terms of a Hermitian generator ${\hat G}_n$
by (\IUndef), and is responsible for transforming the symbols of the
Hamiltonians into functions of the $I^j$ at order $\hbar^n$ (only even
$n$ are used).

The two stages can be summarized schematically by
	\begin{equation}
	\{ {\hat H}^j \} \to \{ {\hat K}^j \} \to
	\{ {\hat M}^j \}.
	\label{stages}
	\end{equation}
All this is very much as in I, but note that a single unitary operator
$\hat U$ simultaneously transforms all the operators ${\hat H}^j$ into
the operators ${\hat K}^j$, and a single set of operators $\{ {\hat
U}_n \}$ transforms all the operators ${\hat K}^j$ into operators
${\hat M}^j$.

\subsection{The preparatory transformation}

The multidimensional method for finding $\hat U$ and the operators
${\hat K}^j$ is similar to the one-dimensional case.  First we
assume that ${\hat U}$ is given, and we work out expressions for the
symbol $K^j$ of the transformed Hamiltonians ${\hat K}^j$.  Then we
impose conditions on these symbols (namely, that the principal symbols
$K^j_0$ should be functions of the $I^j$).  We assume that ${\hat U}$
can be imbedded in a smooth family ${\hat U}_\epsilon$ of unitary
operators with boundary values at $\epsilon=0$ and $\epsilon=1$ as
shown in Table~\ref{table}.  The operators ${\hat U}$ of interest to
us will satisfy this condition.  Next we define the Hermitian operator
${\hat G}_\epsilon$ by (\IGepsilondef), exactly as in I.  The symbol
$G_\epsilon$ of operator ${\hat G}_\epsilon$ of interest to us will be
constructed out of the classical mechanics associated with the
integrable system defined by the principal symbols $H^j_0$.  Since
these have no $\hbar$ dependence, neither does $G_\epsilon$ (unlike
the case of I).

\begin{table}

\caption{\label{table}Notation for operators, symbols and functions
depending on $\epsilon$.}

\begin{indented}
\item[]\begin{tabular}{@{}c|c|cc|c|ccc|cc}
\br %\\
%\mr
$\epsilon=0$ & $1$ & $K^j$ & $K^j_n$ 
	& $\hbox{\rm Id}^\mu$ & $\theta^j$ & $I^j$ & $E^\mu$ & 
	$C^j$ & $C^j_n$ \\
any $\epsilon$ & ${\hat U}_\epsilon$ & $H^j_\epsilon$ & 
	$H^j_{n\epsilon}$ 
	& $Z^\mu_\epsilon$ & $\phi^j_\epsilon$ & $A^j_\epsilon$ 
	& $D^\mu_\epsilon$ 
	& $B^j_\epsilon$ & $B^j_{n \epsilon}$ \\
$\epsilon=1$ & ${\hat U}$ & $H^j$ & $H^j_n$ & 
	$Z^\mu$ & $\phi^j$ & $A^j$ & $D^\mu$
	& $B^j$ & $B^j_n$ \\
\br
\end{tabular}
\end{indented}
\end{table}

Next, as in (\IHepsilondef), we define
	\begin{equation}
	{\hat H}^j_\epsilon = {\hat U}^\dagger_\epsilon {\hat K}^j
	{\hat U}_\epsilon,
	\label{Hhatjepsilondef}
	\end{equation}
so that ${\hat H}^j_\epsilon$ has values at $\epsilon=0,1$ indicated
in Table~\ref{table}.  Differentiating (\ref{Hhatjepsilondef}), we
obtain a generalization of (\IHepsilonode),
	\begin{equation}
	\frac{d {\hat H}^j_\epsilon}{d \epsilon} =
	\frac{i}{\hbar} [ {\hat G}_\epsilon, {\hat H}^j_\epsilon].
	\label{Hhatjepsilonode}
	\end{equation}
As in I, this equation is solved by transcribing it to symbols and
expanding the solution $H^j_\epsilon$ in powers of $\hbar$ (see
(\IHsubnepsilondef)),
	\begin{equation}
	H^j_\epsilon = H^j_{0\epsilon} + \hbar^2 H^j_{2\epsilon}
	+ \hbar^4 H^j_{4\epsilon} + \ldots,
	\label{Hjepsilonexpansion}
	\end{equation}
whereupon we obtain a set of coupled differential equations,
	\numparts
	\begin{eqnarray}
	\frac{D H^j_{0\epsilon}}{D\epsilon} &=0,
	\label{Hsub0epsilonode} \\
	\frac{D H^j_{2\epsilon}}{D\epsilon} &=
	\frac{1}{24} \{G_\epsilon,H^j_{0\epsilon}\}_3,
	\label{Hsub2epsilonode} \\
	\frac{D H^j_{4\epsilon}}{D\epsilon} &=
	\frac{1}{24}\{G_\epsilon,H^j_{2\epsilon}\}_3 -
	\frac{1}{1920}\{G_\epsilon,H^j_{0\epsilon}\}_5,
	\label{Hsub4epsilonode}
	\end{eqnarray}
	\endnumparts
etc., exactly as in (\IHsubepsilonode) except for the $j$
superscripts.  The bracket notation is explained in \IMoyalnotation,
and the notation $D/D\epsilon$ is explained in
Sec.~\Inotationepsilonderiv.  Similarly expanding the symbols $K^j$,
	\begin{equation}
	K^j = K^j_0 + \hbar^2 K^j_2 + \hbar^4 K^j_4 + \ldots,
	\label{Kjexpansion}
	\end{equation}
we find that the functions $H^j_{n\epsilon}$ in
(\ref{Hjepsilonexpansion}) are the solutions of the above differential
equations subject to the initial conditions $H^j_{n\epsilon} = K^j_n$
at $\epsilon=0$.  This implies $H^j_{n\epsilon} = H^j_n$ at
$\epsilon=1$, as shown in Table~\ref{table}.  The boundary conditions
at $\epsilon=1$ differ from those in I, because of the
manner in which the $\hbar$ dependence is handled.

The symbol $G_\epsilon$ is related to an $\epsilon$-dependent
canonical transformation (symplectomorphism) $Z_\epsilon : {\mathbb
R}^{2N} \to {\mathbb R}^{2N}$ with components $Z^\mu_\epsilon :
{\mathbb R}^{2N} \to {\mathbb R}$ (see \Ifunctionnotation\ for
notation), exactly as explained in Sec.~\Icanonicaltransformations.
As in I, when $\epsilon=1$ we drop the $\epsilon$ subscript and write
simply $Z$ or $Z^\mu$, as indicated in Table~\ref{table}.  Then the
solution of (\ref{Hsub0epsilonode}) is
	\begin{equation}
	H^j_{0\epsilon} = K^j_0 \circ Z_\epsilon,
	\label{Hj0epsilonsolution}
	\end{equation}
or, at $\epsilon=1$,
	\begin{equation}
	H^j_0 = K^j_0 \circ Z, \qquad
	K^j_0 = H^j_0 \circ Z^{-1}.
	\label{Kj0solution}
	\end{equation}
See (\IHsubzeroepssoln) and (\IKzerosoln).

Now we choose $Z$ so that $K^j_0$ is a function of the harmonic
oscillator actions $I^j$.  As in I, this is accomplished by using the
action-angle variables $(\phi,A)$ of the integrable system defined by
the Poisson commuting functions $H^j_0$ and those defined by the set
of harmonic oscillator actions $\{I^j\}$, although the details are
considerably more involved than in the one-dimensional case.  The
construction of the action-angle variables $(\phi,A)$ is basically the
standard one in classical mechanics, but one that for our purposes
needs to be examined in some detail.  This is done in
subsections~\ref{normalformquadratic}--\ref{normalformroi} below.  A
consequence of this analysis is that over the region of interest, the
functions $H^j_0$ are invertible functions of the actions $A^j$,
	\begin{equation}
	H^j_0 = f^j_0 \circ A,
	\label{fj0def}
	\end{equation}
where $f^j_0$ are functions $:{\mathbb R}^N \to {\mathbb R}$, or,
collectively, $f_0:{\mathbb R}^N \to {\mathbb R}^N$.  This is the
obvious generalization of (\Ifzerodef) (the 0-subscript means zeroth
order in $\hbar$, see subsection~\ref{secondstage}).  Since the
mapping $f_0$ is invertible in the region of interest, the frequency
matrix $\omega_{jk}$ of (\ref{omegajkdef}) is nonsingular over this
region.  Another notation for the frequency matrix is
	\begin{equation}
	\omega_{jk} = f^j_{0,k}.
	\label{altomegajkdef}
	\end{equation}
(See \Ifunctionnotation\ for notation.)

As for the harmonic oscillators $I^j$, these are their own actions.
We define $\theta = (\theta^1, \ldots, \theta^N)$ as the corresponding
angles, so that $(\theta^j, I^j)$ is just the usual set of
one-dimensional harmonic oscillator action-angle variables in the
$(x_j,p_j)$ plane.  Then the canonical transformation $Z$ satisfies
	\begin{equation}
	\eqalign{
	A^j &= I^j \circ Z, \\
	\phi^j &= \theta^j \circ Z,}
	\label{Zdef}
	\end{equation}
where for simplicity we have set the matrix $\nu$ discussed in
subsection~\ref{cnftorismall} to the $N \times N$ identity.  There is
no loss of generality in this, since the conventional basis contours
on the $H_0$-tori could be chosen to be the pull-backs of the standard
contours $C^j$ on the $I$-tori.  At the end of the calculation it is
easy to carry out a linear transformation on the actions to restore an
arbitrary choice of basis contours on the $H_0$-tori.
Equation~(\ref{Zdef}) is the obvious generalization of (\IZdef) in the
one-dimensional case.

Now we combine the first half of (\ref{Zdef}) with (\ref{fj0def}) to
obtain
	\begin{equation}
	H_0^j \circ Z^{-1} = K^j_0 = f_0^j \circ I,
	\label{K0fnofI}
	\end{equation}
where $I$ denotes the map $:{\mathbb R}^{2N} \to {\mathbb R}^N$ with
components $I^j$.  This shows that the $K^j_0$ are functions of
the $I^j$, as desired.

Having found $Z$, we imbed it in a family $Z_\epsilon$ as in I, with
values at $\epsilon=0,1$ as shown in Table~\ref{table}, and then we
define intermediate (for any $\epsilon$) action-angle variables by
	\begin{equation}
	\eqalign{
	A^j_\epsilon &= I^j \circ Z_\epsilon, \\
	\phi^j_\epsilon &= \theta^j \circ 
		Z_\epsilon,}
	\label{Aphiepsilondef}
	\end{equation}
as in (\IAphiepsilondef).  We also introduce collective notation for
the various action-angle variables (see also (\ref{Dmudef}),
	\begin{eqnarray}
	D^\mu_\epsilon &= (\phi^1_\epsilon, \ldots, \phi^N_\epsilon,
	         A^1_\epsilon, \ldots, A^N_\epsilon),
	\label{Dmuepsilondef} \\
	E^\mu &= (\theta^1, \ldots, \theta^N, I^1, \ldots, I^N),
	\label{Emudef}
	\end{eqnarray}
as indicated in Table~\ref{table}.  Then combining
(\ref{Aphiepsilondef}) with (\ref{Hj0epsilonsolution}), we have 
	\begin{equation}	
	H^j_{0\epsilon} = f^j_0 \circ A_\epsilon,
	\label{Hj0epsilondef}
	\end{equation}
as in (\IHzeroepsaa).  Then we define $G_\epsilon$ as the solution of
(\IZsubepsilonode), as explained in \IGexists.  The inverse Weyl
transform of $G_\epsilon$ is the operator ${\hat G}_\epsilon$, which
defines ${\hat U}_\epsilon$ as the solution of (\IUhatepsilonode),
with value $\hat U$ at $\epsilon=1$.

This completes the solution of (\ref{Hsub0epsilonode}) and the
construction of other objects needed for the 0-th order part of the
preparatory transformation.  The higher order terms are obtained as in
I, for example, we integrate (\ref{Hsub2epsilonode}) to obtain
	\begin{equation}
	K^j_2 = H^j_2 \circ Z^{-1} -\frac{1}{24}
	\int_0^1 d\epsilon \, \{G_\epsilon, H^j_{0\epsilon}
	\}_3 \circ Z^{-1}_\epsilon,
	\label{Kj2solution}
	\end{equation}
with an extra term compared to (\IKtwosoln) because of the different
boundary conditions at $\epsilon=1$ (see Table~\ref{table}). 

\subsection{The second stage}
\label{secondstage}

The second stage transformation proceeds very much as in I, as
indicated by (\ref{Mjdef}) and (\IUndef).  In particular, writing
	\begin{equation}
	M^j = M^j_0 + \hbar^2 M^j_2 + \hbar^4 M^j_4 + \ldots
	\label{Mjexpansion}
	\end{equation}
for the expansion of the symbol of ${\hat M}^j$, we find the hierarchy
of equations,
	\numparts
	\begin{eqnarray}
	M^j_0 &= K^j_0,  \label{M0eqn} \\
	M^j_2 &= K^j_2 + \{G_2,K^j_0\}, \label{M2eqn} \\
	M^j_4 &= K^j_4 + \{G_2,K^j_2\} 
	-\frac{1}{24} \{G_2,K^j_0\}_3 \nonumber \\
	&\qquad + \frac{1}{2}
	\{G_2,\{G_2,K^j_0\}\} + \{G_4,K^j_0\},
	\label{M4eqn}
	\end{eqnarray}
	\endnumparts
etc., just as in (\IMneqns) except for the $j$ superscripts.  As in I,
we wish to choose the generators $G_n$ so that the $M^j_n$ are
functions only of $I$ (i.e., they are independent of $\theta$).  At
zeroth order we have this already,
	\begin{equation}
	M^j_0 = K^j_0 = f^j_0 \circ I = H^j_0 \circ Z^{-1},
	\label{M0soln}
	\end{equation}
as in (\IMzerosoln).  At second order we obtain $M^j_2$ by projecting
out the averaged part of (\ref{M2eqn}),
	\begin{equation}
	M^j_2 = {\bar K}^j_2,
	\label{M2K2avg}
	\end{equation}
as in (\IMtwoKtwoavg), where the overbar indicates an average over the
angles $\theta$ (a torus average).  Projecting out the oscillatory
part, we obtain a differential equation for $G_2$,
	\begin{equation}
	\frac{\partial G_2}{\partial \theta^k} = 
	-(\omega^{-1})_{kj} \,{\tilde K}^j_2,
	\label{G2eqn}
	\end{equation}
as in (\IGtwoeqn), where the tilde indicates the oscillatory part of a
function and where we employ the same abuse of notation mentioned
beneath (\IMzerosoln).

Unlike the one-dimensional case, (\ref{G2eqn}) is not guaranteed a
solution unless an integrability condition is satisfied.  This
question is discussed in \ref{Gnexistence}, where it is shown that the
generators $G_n$ exist to all (even) orders.  The proof relies
critically on the fact that the original operators ${\hat H}^j$
commute to all orders in $\hbar$.  Without this condition, the normal
form transformation does not exist as a power series in $\hbar$.

From this point it is straightforward to proceed as in I to obtain a
generalization of (\IMtworearrangeavg), an expression for $M^j_2$
containing an $\epsilon$-integral of a third order Moyal bracket.
This integral can be done, but it ends up reproducing some of the work
involved in calculating the symbol of a function of an operator, a
problem that is best handled by the methods of Gracia-Saz (2005).  To
take advantage of those methods we proceed somewhat differently.

Granting that the generators $G_\epsilon$, $G_2$, $G_4$, etc.\ exist,
we use the unitary transformations $\hat U$, ${\hat U}_2$, ${\hat
U}_4$, etc.\ to transform the operators ${\hat I}^j$ into new action
operators we denote by ${\hat B}^j$, reversing the sequence of steps
used to transform ${\hat H}^j$ into ${\hat M}^j$ as shown in
(\ref{Mjdef}) and (\ref{stages}).  That is, we define
	\begin{equation}
	{\hat B}^j = {\hat V}^\dagger \, {\hat I}^j \, {\hat V},
	\label{Bjdef}
	\end{equation}
as in subsection~\Iactionoperators.  We call the ${\hat B}^j$ ``action
operators,'' but we do not denote them ${\hat A}^j$ since by our
conventions the latter would be the Weyl transforms of the classical
actions $A^j$.  The operators ${\hat A}^j$ are unitarily equivalent to
the harmonic oscillator actions ${\hat I}^j$ only through lowest order
in $\hbar$, while the operators ${\hat B}^j$ are equivalent to them to
all orders in $\hbar$.  Consequently, the operators ${\hat B}^j$ have
symbols which are equal to the classical actions $A^j$ at lowest
order, that is, if we write
	\begin{equation}
	B^j = B^j_0 + \hbar^2 B^j_2 + \ldots,
	\label{Bjexpansion}
	\end{equation}
then $B^j_0 = A^j$.  As for the first correction term, we find
	\begin{equation}
	B^j_2 = \left(
	\frac{\partial G_2}{\partial \theta^j}
	+ \frac{1}{24} \int^1_0 d\epsilon \,
	\{G_\epsilon, A^j_\epsilon \}_3 \circ Z^{-1}_\epsilon
	\right) \circ Z,
	\label{Bj2def}
	\end{equation}
where we abuse notation in the first term (see \Ifunctionnotation) by
treating $G_2$ both as a function of $(\theta,I)$ and as a map
$:{\mathbb R}^{2N} \to {\mathbb R}$.  The derivation of (\ref{Bj2def})
is given in subsection~\ref{Bjderivation}.

The transformations we have constructed guarantee that the operators
${\hat M}^j$ are functions of the operators ${\hat I}^j$,
	\begin{equation}
	{\hat M}^j = f^j({\hat I}),
	\label{fjdef}
	\end{equation}
for some functions $f^j:{\mathbb R}^N \to {\mathbb R}$, or,
collectively, $f:{\mathbb R}^N \to {\mathbb R}^N$.
Equation~(\ref{fjdef}) is the normal form for the original operators
${\hat H}^j = {\hat V}^\dagger {\hat M}^j {\hat V}$, and it serves to
define the functions $f^j$.  Conjugating (\ref{fjdef}) by ${\hat V}$
and using (\ref{Mjdef}) and (\ref{Bjdef}) gives
	\begin{equation}
	{\hat H}^j 
	= {\hat V}^\dagger f^j({\hat I}) {\hat V}
	= f^j({\hat V}^\dagger {\hat I} {\hat V})
	= f^j({\hat B}),
	\label{HfofB}
	\end{equation}
an equation which determines the eigenvalues of the ${\hat H}^j$ once
$f^j$ and the eigenvalues of the ${\hat B}^j$ are known.  But by
(\ref{Bjdef}) the latter are the same as the eigenvalues of the ${\hat
I}^j$, namely, $(n_j + 1/2)\hbar$, where $n_j \in {\mathbb N} = \{0,1,
\ldots \}$.  

As for the functions $f^j$, to find them we take the symbols of both
sides of (\ref{HfofB}), using the methods of Gracia-Saz (2005) for the
symbol of $f^j({\hat B})$.  This gives
	\begin{equation}
	H^j = f^j + \hbar^2 \Biggl[
	-\frac{1}{16} f^j_{,kl} 
	\xymatrix@C=1pc{(B^k \ar@<0.5ex>[r] \ar@<-0.5ex>[r]
	& B^l)}
	+\frac{1}{24} f^j_{,klm}
	\xymatrix@C=1pc{(B^k \ar[r] & B^l \ar[r] & B^m) }
	\Biggr] + O(\hbar^4),
	\label{HBFoSoOexpn}
	\end{equation}
where $f^j$ and its derivatives are evaluated at $B$ (more properly,
they stand for $f^j \circ B$, etc.)  Then we expand $H$ and $B$ as in
(\ref{Hjhbarexpansion}) and (\ref{Bjexpansion}) and similarly write
$f^j = f^j_0 +\hbar^2 f^j_2 + \ldots$, expanding (\ref{HBFoSoOexpn})
and collecting terms by powers of $\hbar$.  At lowest order we have
$H^j_0 = f^j_0(A)$, showing that $f^j_0$ is the classical
action-energy map (it is the same function $f^j_0$ introduced in
(\ref{fj0def})).  At second order we find
	\begin{equation}
	H^j_2 = f^j_2 + f^j_{0,k} \,B^k_2
	-\frac{1}{16} f^j_{0,kl}
	\xymatrix@C=1pc{(A^k \ar@<0.5ex>[r] \ar@<-0.5ex>[r]
	& A^l)}
	+\frac{1}{24} f^j_{0,klm} 
	\xymatrix@C=1pc{(A^k \ar[r] & A^l \ar[r] & A^m) },
	\label{Hj2def}
	\end{equation}
where now $f^j_n$ and their derivatives are evaluated at $A$.  We
consider everything in this result to be functions of $(\phi,A)$ and
we take the average of both sides to obtain
	\begin{eqnarray}
	f^j_2 &= \Biggl< H^j_2 -f^j_{0,k} \, B^k_2
	+\frac{1}{16} f^j_{0,kl}
	\xymatrix@C=1pc{(A^k \ar@<0.5ex>[r] \ar@<-0.5ex>[r]
	& A^l)} \nonumber \\
	&\qquad -\frac{1}{24} f^j_{0,klm}  
	\xymatrix@C=1pc{(A^k \ar[r] & A^l \ar[r] & A^m) }
	\Biggr>_\phi,
	\label{fj2def}
	\end{eqnarray}
where the angle brackets indicate an average over $\phi$ as in
(\ref{theresult}).  

Equation~(\ref{fj2def}) gives $f^j_2$ in terms of functions that have
been determined, but it would be more satisfactory if the
$\epsilon$-integral contained in the term $B^k_2$ were done (see
(\ref{Bj2def})).  Notice that for the computation of $f^j_2$ (and
hence the quantization rule) we only need the $\phi$-average of
$B^k_2$.  This average kills the first term in (\ref{Bj2def}), since
	\begin{equation}
	\left< \frac{\partial G_2}{\partial \theta^k}
	\circ Z \right>_\phi =
	\left< \frac{\partial G_2}{\partial \theta^k} \right>_\theta
	\circ Z = 0.
	\label{B21sttermvanishes}
	\end{equation}
See (\ref{Zdef}) and \Inotationaveraging.  As for the
$\epsilon$-integral (the second term in (\ref{Bj2def})), it can be
evaluated as follows:
	\begin{equation}
	\int_0^1 d\epsilon \, 
	\xymatrix@C=1pc{(G_\epsilon \ar@<1.0ex>[r]
	\ar[r] \ar@<-1.0ex>[r] & A^j_\epsilon)}
	\circ Z^{-1}_\epsilon \circ Z =
	\frac{1}{2} 
	\xymatrix@C=1pc{D^\mu \ar[r] & A^j
	\ar@<-0.5ex>[r] \ar@<0.5ex>[r] & D_\mu}
	+\frac{\partial X}{\partial \phi^j},
	\label{indefintegral}
	\end{equation}
where $X$ is a function that is not needed for the evaluation of
$f^j_2$ or the quantization condition, since its $\phi$-derivative
vanishes upon $\phi$-averaging.  This result is derived in
\ref{antiderivative}.  Thus, we have
	\begin{equation}
	\left< B^k_2 \right>_\phi = \frac{1}{2}
	\left< 
	\xymatrix@C=1pc{D^\mu \ar[r] & A^j
	\ar@<-0.5ex>[r] \ar@<0.5ex>[r] & D_\mu}
	\right>_\phi.
	\label{Bk2avg}
	\end{equation}

Using now (\ref{Bk2avg}) in (\ref{fj2def}) and writing $\omega_{jk} =
f^j_{0,k} \circ A$ (see (\ref{omegajkdef}) and (\ref{altomegajkdef})),
we obtain the second order term of (\ref{theresult}),  apart
from the special choice of contours $\Gamma^j$, which forces the
ranges of the quantum numbers $n_j$ and the Maslov indices to take on
their values for the harmonic oscillator ($n_j \in {\mathbb N}$ and
$\gamma_j=2$).  But the entire expression is covariant under a
$GL(N,{\mathbb Z})$ transformation of actions and angles, so if we set
$A' = \nu A$, $\phi' = \nu^{-1} \phi$ where $\nu \in GL(N,{\mathbb
Z})$, then we have $n'_j = \sum_k \nu_{jk} \, n_k$ and $\gamma'_j = 2
\sum_k \nu_{jk}$.  Dropping the primes, we then have the general
formula (\ref{theresult}).  Further discussion of this result is
presented in Sec.~\ref{discussion}.

\section{Details of the calculation}
\label{details}

In this section we present a series of digressions from the overview
of Sec.~\ref{thecalculation} that supply details of the calculation.  
 
\subsection{Classical normal form in a small neighborhood of $z_0$}
\label{normalformquadratic}

In this subsection we develop the classical normal form transformation
$Z$ in a small neighborhood of the fixed point $z_0$ of the
Hamiltonian $H^1_0$, in which the functions $H^j_0$ are well
approximated by their quadratic expansions about $z_0$.  (Properly,
this discussion should be framed in terms of tangent spaces and germs
of smooth functions, but we shall proceed somewhat intuitively.)  At
this level of description, $Z$ is a linear canonical transformation
(the linearization of the fully nonlinear $Z$ defined over the region
of interest).  To begin we list the assumptions we make about the
behavior of the principal symbols $H^j_0$ in a neighborhood of the
fixed point $z_0$.  These amount to taking the most generic case of a
stable fixed point of an integrable system.

We assume that all symbols $H^j$ (that is, all the terms $H^j_n$ of
their $\hbar$ expansions) are smooth in a neighborhood of $z_0$.  This
excludes the case of singular potentials, such as the Coulomb
potential.  We define $Q^j$ as the $2N \times 2N$ real, symmetric
Hessian matrix of the principal symbol $H^j_0$ evaluated at $z_0$,
	\begin{equation}
	Q^j_{\mu\nu} = \frac{\partial^2 H^j_0}
	{\partial z^\mu \partial z^\nu}(z_0).
	\label{Qjdef}
	\end{equation}
We assume that $Q^1$ (the Hessian of the Hamiltonian) is positive or
negative definite, corresponding to stable motion about the fixed
point.  For convenience, if $Q^1$ is negative definite, we agree to
replace ${\hat H}^1$ by $-{\hat H}^1$ at the beginning of the
calculation, so that in the following we can assume that $Q^1$ is
positive definite.  The $Q^j$ for $j>1$ need not be definite (positive
or negative).  We assume that the set of matrices $\{ Q^j, j=1,
\ldots, N\}$ is linearly independent.    For simplicity, we assume
that all the principal symbols $H^j_0$ vanish at $z_0$ (if not, we
just subtract constants from the operators ${\hat H}^j$ to make this
so). 

These assumptions imply that the first derivatives of the $H^j_0$ at
the fixed point, denoted by
	\begin{equation}
	V^j_\mu = \frac{\partial H^j_0}{\partial z^\mu} (z_0),
	\label{Vjmudef}
	\end{equation}
all vanish.  For $j=1$, this follows from the assumption that $z_0$ is
a fixed point of $H^1_0$.  For $j>1$, we take the derivative of the
vanishing Poisson bracket $\{ H^1_0, H^j_0 \}$ and evaluate it at
$z_0$:
	\begin{equation}
	0 = \left.\frac{\partial}{\partial z^\mu}
	\{ H^1_0, H^j_0 \}\right|_{z_0}
	= Q^1_{\mu\alpha} J^{\alpha\beta} V^j_\beta +
	V^1_\alpha J^{\alpha\beta} Q^j_{\beta\mu}.
	\label{Vjvanishes}
	\end{equation}
The second term on the right hand side vanishes since $V^1_\mu=0$,
which implies that $V^j_\mu=0$ for $j>1$, since both $Q^1_{\mu\nu}$
and $J^{\mu\nu}$ are nonsingular matrices (in the former case, because
$Q^1$ is positive definite).  Thus, the Hamiltonians $H^j_0$ all have
a common fixed point at $z_0$, and their Taylor series expansion
through quadratic order is
	\begin{equation}
	H^j_0(z) = \frac{1}{2} Q^j_{\mu\nu} 
	(z-z_0)^\mu (z-z_0)^\nu.
	\label{Hj0expansion}
	\end{equation}

Now computing the second derivatives of the vanishing Poisson bracket
$\{ H^j_0, H^k_0 \}$ and evaluating at $z_0$, we obtain the matrix
equations,
	\begin{equation}
	[JQ^j, JQ^k] = 0,
	\label{QjQkcommutator}
	\end{equation}
or
	\begin{equation}
	Q^j J Q^k = Q^k J Q^j = -(Q^j J Q^k)^T = {\rm antisymmetric},
	\label{altQjQkcommutator}
	\end{equation}
where $T$ indicates the transpose.  

Regarding the set of matrices $\{ Q^j \}$, we have the following
theorem.  Write the harmonic oscillator actions in the form,
	\begin{equation}
	I^j(z) = \frac{1}{2} (x_j^2 + p_j^2)
	= \frac{1}{2} z^T \Delta^j z,
	\label{Deltajdef}
	\end{equation}
which defines $\Delta^j$ is the $2N \times 2N$ symmetric matrix that
contains zeros everywhere except for ones in the $(j,j)$ and
$(N+j,N+j)$ positions.  Then there exists a symplectic matrix $S$ such
that
	\begin{equation}
	S^T Q^j S = \amatrix_{jk}\, \Delta^k,
	\label{matrixnormalform}
	\end{equation}
where $\amatrix_{jk}$ is a real, $N \times N$, nonsingular matrix.
Moreover, the numbers $\amatrix_{1k}$ (the first row of the matrix)
are all positive.  The proof of this theorem is discussed in
\ref{quadraticnormalform}.  The proof relies on the fact that $Q^1$ is
positive definite. 

Now we define a canonical transformation $Z$ in terms of the symplectic
matrix $S$ produced by this theorem,
	\begin{equation}
	Z(z) = S^{-1}(z-z_0).
	\label{Zdeflinearized}
	\end{equation}
This allows us to combine (\ref{Hj0expansion}) and
(\ref{matrixnormalform}) to give 
	\begin{equation}
	H^j_0 = \amatrix_{jk} \, I^k \circ Z.
	\label{linearnormalform}
	\end{equation}
Thus, under the transformation $Z$ (or $Z^{-1}$), the Hamiltonians
$H^j_0$ are mapped into functions of the harmonic oscillators
$I^j$ (in fact, linear functions).  Therefore we identify this $Z$
with the transformation $Z$ used in (\ref{Kj0solution}), and we have
the desired normal form transformation in a small neighborhood of the
fixed point,
	\begin{equation}
	K^j_0 = \amatrix_{jk} \, I^k.
	\label{K0linearsolution}
	\end{equation}

\subsection{Tori, contours and actions in a small neighborhood of
$z_0$}
\label{cnftorismall}

The level sets of the harmonic oscillator actions $I^j$ (say,
$I^j=c_j$, for $c_j \ge 0$) are $n$-tori, where $n$ is the number of
the $c_j$ that do not vanish.  Thus, $0 \le n \le N$, and the subset
of ${\mathbb R}^{2N}$ that contains tori of dimension $\le n$ is the
union of a set of $N!/n!(N-n)!$ planes of dimension $2n$ passing
through the origin.  On the other hand, (\ref{linearnormalform}) shows
that $Z$ maps the level sets of the functions $H^j_0$ near $z_0$ onto
those of the harmonic oscillators $I^j$ near the origin.  Thus, the
level sets of the $H^j_0$ near $z_0$ are also tori of dimension
ranging from 0 to $N$ that fit together exactly like those of the
harmonic oscillators $I^j$ near the origin.  We shall call these two
sets of tori the ``$H_0$-tori'' and the ``$I$-tori.''

To visualize how the tori of various dimension fit together, we
consider the two maps,
	\begin{equation}
	\eqalign{
	H_0:{\mathbb R}^{2N} &\to {\mathbb R}^N : 
	z \mapsto \bigl(H^1_0(z), \ldots, H^N_0(z)\bigr) \\
	I:{\mathbb R}^{2N} &\to {\mathbb R}^N :
	z \mapsto \bigl(I^1(z), \ldots, I^N(z)\bigr)}
	\label{COMmaps}
	\end{equation}
where ${\mathbb R}^N$ can be thought of as ``constant-of-motion
space.''  These are examples of the energy-momentum map discussed by
Cushman and Bates (1997).  We call these the ``$H_0$-map'' and the
``$I$-map,'' respectively.  These maps and others associated with them
are illustrated in Fig.~\ref{EMmapsinfinitesimal}.  An entire level
set of the functions $H^j_0$ or $I^j$ is mapped by these two maps into
a single point of constant-of-motion space.  In the following we think
of the domain of the $H_0$-map as a small region $U$ around $z_0$ (an
open ball).  The domain of the $I$-map may be taken to be all of
${\mathbb R}^{2N}$, but we shall often be interested in the small
region $V=Z(U)$ around the origin (so $V$ is also an open ball).  The
images of these maps are the sets of values the constants of motion,
$H^j_0$ or $I^j$, can take on.

The image of ${\mathbb R}^{2N}$ under the $I$-map is the region $c_j
\ge 0$, for $j=1, \ldots, N$, for example, it is a single octant of
${\mathbb R}^3$ when $N=3$.  Thus, the image of $V$ under the $I$-map
looks like a corner of an $N$-cube that has been sliced off in such a
manner that the cut does not pass through the vertex itself, as
illustrated by the shaded area in the upper right of
Fig.~\ref{EMmapsinfinitesimal}.  The sliced corner has flat faces of
dimensionality $n$, ranging from 0 to $N$, where the $n=0$ face is the
vertex (the origin of ${\mathbb R}^N$), and the $n=N$ face is the
entire sliced corner.  In the figure, the 1-faces are heavy lines,
coinciding with the positive axes.  The inverse image of a point in
the interior of an $n$-face of the sliced corner is a torus in
${\mathbb R}^{2N}$ of dimensionality $n$.

By throwing out all the faces with $n<N$, we can create an open,
contractible region of ${\mathbb R}^N$, whose inverse image under the
$I$-map is the set of invariant $N$-tori of the harmonic oscillators.
This set is topologically the Cartesian product of the just mentioned
region and an $N$-torus.  Thus, the subset of phase space ${\mathbb
R}^{2N}$ containing the $N$-tori is connected and has the form of a
trivial fiber bundle in which the fiber is an $N$-torus.

Now let $H^j_0 = E^j$ for $j=1, \ldots, N$ be a level set of the
$H^j_0$ near $z_0$, which is mapped by $Z$ onto a level set $I^j =
c_j$ near the origin.  Then by (\ref{linearnormalform}) we have
	\begin{equation}
	E^j = \amatrix_{jk} \, c_k,
	\label{levelsetmap}
	\end{equation}
which means that the image of $U$ under the $H_0$-map is just a linear
transformation (specified by $\amatrix_{jk}$) applied to the image of
$V$ under the $I$-map (the sliced corner of the $N$-cube).  In
particular, all the level sets of the functions $H^j_0$ that are
$N$-tori correspond to interior points of the sliced corner of the
$N$-cube, while tori of lower dimension reside on the surface.  Thus,
the image of $U$ under the $H_0$-map has a ``corner.''  See, for
example, Fig.~3.3 in Chapter~IV in Cushman and Bates (1997) (in that
figure, the corner does not occur at the origin of ${\mathbb R}^N$,
because the constants of motion do not take on the value 0 at the
fixed point).

Now we consider basis contours on the various tori.  A set of closed
contours on a torus constitutes a basis if any closed contour can be
represented as a linear combination of the basis contours with integer
coefficients, with the understanding that two closed contours are
considered equal if they can be continuously deformed into one
another.  That is, the contours form a basis of the first homology
group $H^1({\mathbb Z})$ of the torus, with homologous contours
considered equal.  On a specific $N$-torus, the choices of basis
contours are related by $GL(N,{\mathbb Z})$ transformations (see
\ref{notation}).

As for the $I$-tori, the functions $I^j(z)$ are themselves actions of
these tori, that is, there exist basis contours $\{ C^j \}$ on the
$I$-tori such that
	\begin{equation}
	I^j(z) = \frac{1}{2\pi} \oint_{C^j} p_k \, dx_k.
	\label{Ijintegrals}
	\end{equation}
Specifically, we shall take the contour $C^j$ to be the circle in the
$x_j$-$p_j$ plane whose radius is determined by the contour value of
$I^j$, with the other $x_k$ and $p_k$ fixed.  This gives a smooth
assignment of contours $C^j$ on the $I$-tori over all of ${\mathbb
R}^{2N}$, and hence over the region $V$.  

As for basis contours on the $H_0$-tori, we could take these to be
$Z^{-1} \circ C^j$, where we think of the $C^j$ as maps from the
circle onto phase space, $C^j: S^1 \to {\mathbb R}^{2N}$, but in
practice other basis contours on the $H_0$-tori, call them $\Gamma^j$,
may be more convenient (for example, see \ref{normalformexample}).  On
each $H_0$-torus, the contours $\Gamma^j$ must be related to $Z \circ
C^j$ by an element $\nu \in GL(N,{\mathbb Z})$, so that
	\begin{equation}
	Z \circ \Gamma^j = \nu_{jk} \, C^k.
	\label{Zcontourmap}
	\end{equation}
By requiring that $\nu$ be the same for all $H_0$-tori, the contours
$\Gamma^j$ constitute a smooth assignment of basis contours over the
$H_0$-tori in the small region $U$.  

Once the $\Gamma^j$ are given, the actions $A^j$ are defined for $z
\in U$
	\begin{equation}
	A^j(z) = \frac{1}{2\pi} \oint_{\Gamma^j} p_k \, dx_k,
	\label{Ajdef}
	\end{equation}
and they are smooth functions over $U$.  Then since the symplectic
2-form $d(p_k \, dx_k)$ is invariant under the canonical
transformation $Z$, we can map the integral (\ref{Ajdef}) from the
$H_0$-torus to the $I$-torus to obtain
	\begin{equation}
	A^j = \nu_{jk}\, I^k \circ Z,
	\label{AjlcofIk}
	\end{equation}
which is the first half of (\ref{Zdef}), here established over small
neighborhood $U$ around $z_0$.  

As for the angles, we first define the harmonic oscillator angles
$\theta^j$ as the clockwise angles in the $x_j$-$p_j$ plane, vanishing
on the positive $p_j$-axis (a convenient definition).  Then the second
half of (\ref{Zdef}) can be taken as the definition of the angles
$\phi^j$ of the $H_0$-tori.

Finally we combine (\ref{linearnormalform}) and (\ref{AjlcofIk}) to
obtain, 
	\begin{equation}
	H^j_0 = (\amatrix \nu^{-1})_{jk} \, A^k,
	\label{HofAlinearized}
	\end{equation}
which gives the form of the functions $H^j(A)$ in neighborhood $U$.
Differentiating this and using (\ref{omegajkdef}), we obtain
	\begin{equation}
	\omega_{jk}(0) = (\amatrix \nu^{-1})_{jk},
	\label{omegaanu}
	\end{equation}
where the frequency matrix is considered a function of the actions
$A$, and is here evaluated at $A=0$ (corresponding to the fixed
point).  This matrix is nonsingular, since both $\amatrix_{jk}$ and
$\nu_{jk}$ are nonsingular.

This completes the classical normal form transformation in a small
neighborhood of the fixed point $z_0$.  \ref{normalformexample}
contains a physical example of this transformation, in which all the
constructions of this subsection are carried out explicitly.

\subsection{Classical normal form in the region of interest}
\label{normalformroi} 

In this subsection we discuss the definition of the ``region of
interest'', denoted $R$, an enlargement of the small neighborhood $U$
of the fixed point $z_0$ which is the domain of the (generally
nonlinear) canonical transformation $Z$ that brings about the
classical normal form (\ref{Kj0solution}) and satisfies (\ref{Zdef}).
The situation is illustrated in Fig.~\ref{mapsroi}, which is a
modified version of Fig.~\ref{EMmapsinfinitesimal}.  The region $R$
and the fixed point $z_0$ are shown in the lower left corner of this
figure.  The transformation $Z$ maps $R$ onto the region $W=Z(R)$,
upon which the harmonic oscillator actions $I$ are defined.  The
mappings $H_0$ and $I$ are the same in Fig.~\ref{mapsroi} as in
Fig.~\ref{EMmapsinfinitesimal}, except for the expansion of the
domains.  Note that in subsection~\ref{normalformquadratic}, the notation
$Z$ stood for the linearization of the $Z$ that we are discussing in
this subsection.  For example, (\ref{Zdeflinearized}) must now be
understood as
	\begin{equation}
	\frac{\partial Z^\mu}{\partial z^\nu} (z_0) = 
	(S^{-1})^\mu{}_\nu,
	\label{Zlinearized}
	\end{equation}
with $Z(z_0)=0$.

A simple approach to understanding the transformation $Z$ is just to
say that it maps the action-angle variables of the Hamiltonians
$H_0^j$ into those of the harmonic oscillators $I^j$, as in
(\ref{Zdef}). This approach requires the action-angle variables for
$H^j_0$ to be defined.  Then $Z$ is canonical because it is the
composition of two canonical transformations.  Moreover it is easy to
see that $Z$ is smooth, except possibly where the $N$-tori degenerate
into tori of lower dimension and some of the angles become undefined.
In I worries about the smoothness of $Z$ at the fixed point led us
to take another approach, in which the Morse lemma was used first to
construct $Z$, and then $Z$ was used to define the angles $\phi$ (see
\IZissmooth). Nevertheless, the simpler approach is still useful, in 
that it shows that the domain of $Z$ in the one-dimensional problem
can be extended out from the fixed point to the first separatrix,
where the action-angle variables become undefined.

In the multidimensional case, the construction of $Z$ can again be
based on symplectic generalizations of the Morse lemma.  This has been
done by Eliasson (1990), in more generality than we need here.  In the
following we shall simply assume the existence of a smooth, symplectic
map $Z:R \to {\mathbb R}^{2N}$ such that $Z(z_0)=0$ and such that
$H_0$-tori are mapped onto $I$-tori.  As in I, we shall take $R$ to be
the largest region containing $z_0$ over which the entire construction
works; $R$ is required to be a union of invariant $H_0$-tori. In the
multidimensional case, however, obstacles to the extension of the
domain $R$ include not only separatrices but also pinched tori and
other kinds of nongeneric level sets of the functions $H^j_0$.  The
following remarks are useful for visualizing the kinds of such level
sets that can arise in the multidimensional case.

First we comment on the Liouville-Arnold theorem, which says that if
the differentials $dH^j_0$ are linearly independent on a
compact, connected component of a level set of the Poisson commuting
functions $H^j_0$, then that level set is a Lagrangian $N$-torus.  By
the assumptions we have made, the levels sets of the $H^j_0$ are
certainly compact in a sufficiently small neighborhood of $z_0$,
because the level sets of $H^1_0$ alone are compact there (they are
the spheres $S^{2N-1}$).  But the differentials $dH^j_0$ are not
necessarily linearly independent on those level sets, even in a small
neighborhood of $z_0$, as shown by the invariant $n$-tori for $n<N$,
which exist arbitrarily close to $z_0$ in the manner described in
subsection~\ref{cnftorismall}. For example, at the fixed point itself (a
0-torus), the differentials $dH^j_0$ all vanish.

We consider the rank of the set of differentials $\{ dH^j_0 \}$ to be
the rank of the $N \times 2N$ matrix $M_{j\mu} = \partial H^j_0
/\partial z^\mu$.  This rank is constant along any Hamiltonian flow
generated by any of the $H^j_0$, as are the Hamiltonian functions
$H^j_0$ themselves.  The commuting Hamiltonian flows associated with
the $H^j_0$ define an ${\mathbb R}^N$-action on phase space, assuming
that the advance maps exist for all time.  Thus a level set of the
$H^j_0$ consists of a union of orbits of the ${\mathbb R}^N$-action,
upon each of which the rank of the differentials $dH^j_0$ is constant.
Each of these orbits is diffeomorphic to ${\mathbb R}^r
\times (T^1)^s$, where $r+s=n$ (a generalized cylinder), where $T^1$
is the 1-torus (a circle), and where the dimension of the orbit $n$
equals the rank of the differentials $dH^j_0$.

The differentials $dH^j_0$ can fail to be linearly independent on a
level set in different ways.  The $n$-tori for $n<N$ that exist even
near the fixed point are composed of a single orbit of the ${\mathbb
R}^N$-action, upon which the rank of the differentials is $n$.  The
separatrix in a double well oscillator in one dimension consists of
three orbits of the ${\mathbb R}^1$-action, the two orbits that
asymptote to the fixed point as $t \to \pm \infty$, each diffeomorphic
to ${\mathbb R}$, and the fixed point itself, a 0-torus.
Generalizations of this behavior occur in higher dimensions.  Another
example is offered by the pinched torus of the spherical pendulum,
which exhibits the phenomenon of monodromy (Duistermaat 1980, Cushman
and Duistermaat 1988, Cushman and Bates 1997, Alber \etal 1998, Child
1998, Cushman \etal 2004).  In this case the pinched torus minus the
pinch point is a 2-dimensional surface diffeomorphic to ${\mathbb R}
\times S^1$, while the pinch point itself is a 0-torus.

Another issue concerns the definitions of the actions $A^j$ as smooth
functions $:R \to {\mathbb R}^N$, which requires a smooth assignment
of basis contours over the $H_0$-tori in $R$.  Consider a region of
phase space in which the differentials $dH^j_0$ are linearly
independent and the level sets of the $H^j_0$ are $N$-tori, and then
consider the image of that region under the $H_0$-map.  This image can
be taken to be the base space of a discrete, principal fiber bundle,
in which the fiber over a point is the set of all basis contours on
the corresponding $N$-torus.  The structure group of this bundle is
$GL(N,{\mathbb Z})$, which relates the sets of basis contours.  A
smooth assignment of basis contours as a field over the base space is
equivalent to a global section of this bundle, which exists if and
only if the bundle is trivial.  But such bundles are nontrivial
precisely where monodromy occurs (this is the meaning of monodromy).
Notice that if we have a smooth $Z$ that maps $H_0$-tori in some
region onto $I$-tori, we can pull back the basis contours of the
$I$-tori, which can be smoothly assigned, to create a smooth
assignment of basis contours on the $H_0$-tori.  Thus, monodromy
cannot occur in the domain of any such $Z$, such as the region $R$
considered here.  In the example of monodromy given in Fig.~3.3,
Chapter IV of Cushman and Bates (1997), one would have to introduce a
branch cut terminating on the pinch point in order to create a region
$R$ free of monodromy.  There is no unique way to do this, and hence
no unique, maximal set $R$ upon which $Z$ is defined.

Finally, we impose the requirement that $R$ be contractible, since
this guarantees the existence of the generator $G_\epsilon$ (see
\IGexists).  Note that this condition by itself guarantees that any 
bundle over $R$ is trivial.

To proceed we assume $Z$ is defined on some suitable set $R$, and that
its linearization at $z_0$ agrees with the transformation previously
worked out in subsection~\ref{normalformquadratic}. A smooth
assignment of basis contours $\Gamma^j$ on the $H_0$-tori in $R$ is
made according to (\ref{Zcontourmap}), where $\nu$ is the same matrix
constructed in the small neighborhood of $z_0$ (it is extended to all
of $R$).  Then (\ref{AjlcofIk}) holds in all of $R$, and maps $A$ and
$\nu$ in Fig.~\ref{mapsroi} are defined and satisfy $A = \nu
\circ I \circ Z$.  One nonobvious conclusion from this is that the
boundaries of the image of the $A$-map are flat (straight lines in
Fig.~\ref{mapsroi}).   Another conclusion is that the actions $A^j$
uniquely label the $H_0$-tori, that is, no two $H_0$-tori in $R$ have
the same actions.  This follows since the actions $I^j$ uniquely label
the $I$-tori. 

This means that the map $f_0$ in Fig.~\ref{mapsroi} is defined, for if
we take a point in the image of the $A$-map, its inverse image is an
invariant $H_0$-torus, upon which the values of the $H^j_0$ are
defined.  Thus, $H_0 = f_0 \circ A$ as in (\ref{fj0def}).  Moreover,
the map $f_0$ is of maximal rank ($N$) and therefore invertible in the
region containing $N$-tori, where the differentials $dH^j_0$ are
linearly independent.  This means that the frequency matrix
$\omega_{jk}(A)$ is invertible in this region.  The frequency matrix
can be defined on the boundaries of the image of the $A$-map by a
limiting process; this limit is nonsingular at the fixed point (see
(\ref{omegaanu})).  We shall assume that the limit at other points on
the boundary is also nonsingular (this excludes the possibility that
the kind of nongeneric behavior seen in the quartic oscillator in one
dimension might develop as we move along the space of tori of
dimension $n<N$).  Thus, the image of the $H_0$-map is a smooth
deformation of the truncated corner of the $N$-cube, as illustrated in
Fig.~\ref{mapsroi} (with corners and curved faces).  These features
are clear in the example of Fig.~3.3, Chapter IV of Cushman and Bates
(1997).  Finally, we define the map $K_0$ as $H_0 \circ Z^{-1}$, as in
(\ref{Kj0solution}); this implies $K_0 = f_0 \circ \nu \circ I$, as in
(\ref{K0fnofI}) (where we have replaced $\nu$ by the identity).

\subsection{Symbols $B^j$ of the action operators ${\hat B}^j$}
\label{Bjderivation}

The derivation of the symbols $B^j$ of the action operators ${\hat
B}^j$, defined by (\ref{Bjdef}), is a straightforward application of the
method used to derive the symbols $M^j$ from the symbols $H^j$,
except that it is run backwards.  For convenience we define an
intermediate set of operators ${\hat C}^j$ by ${\hat B}^j = {\hat
U}^\dagger \, {\hat C^j} \, {\hat U}$, that is,
	\begin{equation}
	{\hat C}^j = {\hat U}^\dagger_2 {\hat U}^\dagger_4 \ldots
	{\hat I}^j \ldots {\hat U}_4 {\hat U}_2.
	\label{Cjhatdef}
	\end{equation}
Then by expanding the ${\hat U}_n$ in terms of their generators and
transcribing to symbols, we obtain
	\begin{equation}
	C^j = I^j + \hbar^2 \{ G_2, I^j \} + \hbar^4 \left[
	\frac{1}{2} \{ G_2, \{ G_2, I^j \}\}
	+ \{ G_4, I^j \} \right] + O(\hbar^6),
	\label{Cjdef}
	\end{equation}
where third and higher order Moyal brackets involving $I^j$ are
dropped since $I^j$ is a quadratic polynomial.

Next we define ${\hat B}^j_\epsilon = {\hat U}^\dagger_\epsilon \,
{\hat C}^j {\hat U}_\epsilon$, take $\epsilon$-derivatives to obtain a
differential equation for the ${\hat B}^j_\epsilon$, and transcribe
them to symbols, exactly as in the steps leading up to
(\ref{Hsub0epsilonode})--(\ref{Hsub4epsilonode}).  The
$\hbar$-expansion of the symbols $B^j_\epsilon$ yield terms $B^j_{n
\epsilon}$ with boundary values at $\epsilon=0,1$ shown in
Table~\ref{table}.  Note that $C^j_0 = I^j$.  The differential
equations are easily solved, yielding $B^j_0 = A^j$ and (\ref{Bj2def})
through second order.

\section{Discussion of the torus quantization rule (\ref{theresult})}
\label{discussion}

In this section we discuss various aspects of the torus quantization
rule (\ref{theresult}), the normal form transformation, and the
calculation used to derive them.

\subsection{Doing the $\epsilon$-integrals}
\label{epsilonintegrals}

In both I and this paper, the $\epsilon$-integrals that arose in the
computation of the normal form function $f$ all turned out to be
doable, at least through second order in $\hbar$, although the methods
employed were {\it ad hoc} and no proof was given that the same could
be done at higher orders.  See \Iantideriv\ and \ref{antiderivative}.
In principle it would be possible to carry out the normal form
transformation using the integral representation of the Moyal star
product, so that the conjugation (\ref{Kjdef}) could be evaluated by
the stationary phase approximation.  This approach would avoid the
necessity of imbedding ${\hat U}$ or $Z$ in a one-parameter family of
transformations, and there would be no $\epsilon$-dependent paths
through various group manifolds to integrate over. (But one would have
to work with the rapidly oscillating symbol $U$ of $\hat U$, something
we have avoided in the approach of this paper.)  The existence of this
alternative approach suggests that all $\epsilon$-integrals that arise
in the present approach that are needed for the normal form are
doable, that is, the integrals can be expressed in terms of the
transformations at the end points of the path (that is, $Z_\epsilon$
evaluated at $\epsilon=0,1$), and do not depend on the path itself.
An $\epsilon$-integral that is independent of path must mean that in
some sense we are integrating an exact 1-form on the
infinite-dimensional group manifold.  See \IuniquenessofK2, in which
the $\epsilon$-integral occurring in the expression for $K_2$ in the
one-dimensional case was shown to change by an exact
$\theta$-derivative upon a variation of the path, that is, exactly
what was needed to guarantee that the normal form be invariant.  The
calculation of the variation of the $\epsilon$-integral in that case
relied on the star version of the Jacobi identity.  It is likely that
a general understanding of the path dependence of the
$\epsilon$-integrals in the general case (multidimensional, arbitrary
order in $\hbar$) would result by following the suggestions raised by
this calculation.

\subsection{The quantization rule involves the angles}
\label{anglesintheresult}

A striking aspect of the quantization rule (\ref{theresult}) is the
presence of the angles angles $\phi^k$ in the first diagram of the
second order term, which we see if we write $D^\mu=(\phi,A)$.  This
diagram is the difficult one derived in \ref{antiderivative}.  The
appearance of angles in the quantization condition requires some
discussion.

First, we note that in the one-dimensional problem the angles can be
eliminated from the quantization condition.  This is shown to second
order in $\hbar$ by the actual result in I, and to fourth order by
Gracia-Saz (2005), who used Colin de Verdi\`ere's (2004)
method for computing the eigenvalue series combined with his own
theory of the symbol of a function of an operator.  In fact, it is
implicit in Gracia-Saz's work that the angles can be eliminated to all
orders in $\hbar$ in the one-dimensional problem.

Next, we note that the angles are not unique, since their origin on
each torus (the point where the angles are taken to vanish) can be
chosen with a great deal of freedom.  That is, given one set of
action-angle variables $(\phi,A)$, it is always possible to subject
them to a canonical transformation of the form,
	\begin{equation}
	\eqalign{
	\phi^{\prime j} &= \phi^j + 
	\frac{\partial F}{\partial A^j}, \\
	A^{\prime j} &= A^j}
	\label{angleoriginshift}
	\end{equation}
where $F(A)$ is some function of the actions.  Geometrically, the
surface $\phi=0$ is required to be a Lagrangian manifold.  The
quantization condition (\ref{theresult}) must be independent of such a
shift in the origin of the angles.

To show that it is, we write the $\phi$-dependent diagram in
(\ref{theresult}) as
	\begin{equation}
	\xymatrix@C=1pc{D^\mu \ar[r] & A^j 
	\ar@<-0.5ex>[r] \ar@<0.5ex>[r] & D_\mu} =
	-\xymatrix@C=1pc{\phi^k \ar[r] & A^j 
	\ar@<-0.5ex>[r] \ar@<0.5ex>[r] & A^k} +
	\xymatrix@C=1pc{A^k \ar[r] & A^j 
	\ar@<-0.5ex>[r] \ar@<0.5ex>[r] & \phi^k}.
	\label{2phidiagrams}
	\end{equation}
Under the replacement $\phi^k \mapsto \phi^k + F_{,k}(A)$, the right
hand side goes into itself plus the sum of three terms, of which two
cancel by simple symmetry considerations.  The remaining term is
	\begin{equation}
	F_{,klm}\vcenter{
	\xymatrix@C=1pc@R=0.3pc{&&A^l \cr
	A^k \ar[r] & A^j \ar[ur] \ar[dr] \\
	& & A^m}},
	\label{3angleterms}
	\end{equation}
which by identity (\ref{wheelidentity}) can be written as an exact
$\phi^j$-derivative,
	\begin{equation}
	F_{,klm} 
	\xymatrix@C=1pc{A^j \ar[r] & (A^k \ar[r] & A^l \ar[r]
	& A^m)} =
	- \frac{\partial}{\partial \phi^j}[ F_{,klm}
	\xymatrix@C=1pc{(A^k \ar[r] & A^l \ar[r]
	& A^m)]},
	\label{anglecorrectionterm}
	\end{equation}
which vanishes upon the $\phi$-averaging in (\ref{theresult}).
Therefore, although the diagram (\ref{2phidiagrams}) itself is not
invariant under the shift of origin (\ref{angleoriginshift}), its
angle average is, and hence so is the quantization condition
(\ref{theresult}).

A third aspect of the appearance of angles in the torus quantization
rule (\ref{theresult}) is that in practice angles are much harder to
compute than actions, either numerically or as explicit functions of
$x$'s and $p$'s, and what is worse is computing the derivatives of the
angles with respect to $z^\mu$.  This makes a numerical test of the
quantization rule (\ref{theresult}) difficult.  Since the quantization
condition itself does not depend on the origin of the angles, one
wonders whether there is a method of computing the second order term
that does not rely on this origin, either.  We do not know the answer,
but the following are some relevant considerations.

First, not all diagrams with angles cause difficulties.  For example,
the angle average of a diagram of the form $\xymatrix@C=1pc{\phi^k
\ar[r] & (X)}$, where $X$ is any diagram, can be written in the form,
	\begin{equation}
	\left< \xymatrix@C=1pc{\phi^k \ar[r] & (X)} \right>_\phi =
	\left< \frac{\partial X}{\partial A^k} \right >_\phi =
	\frac{\partial}{\partial A^k} \left< X \right>_\phi,
	\label{innocuousdiagram}
	\end{equation}
in which the operator $\partial/\partial A^k$ has been taken out of
the average.  This operator, when applied to a function of both $A$
and $\phi$, is not invariant under the transformation
(\ref{angleoriginshift}), but it is so when applied to a function of
$A$ only.  In fact, such action derivatives, outside an averaging
operator, occur in the one-dimensional quantization rule.

One might hope to transform the difficult diagram into another one
that is independent of the origin of the angles, plus terms
that average to zero.  We doubt that this can be done, but the
following identity is a interesting attempt along these lines:
	\begin{eqnarray}
	\xymatrix@C=1pc{D^\mu \ar[r] & A^k \ar@<-0.5ex>[r] 
	\ar@<0.5ex>[r] & D_\mu} =
	\xymatrix@C=1pc{D^\mu \ar[r] & (A^k \ar@<-0.5ex>[r] 
	\ar@<0.5ex>[r] & D_\mu)} +
	\vcenter{
	\xymatrix@C=0pc@R=0.8pc{& A^k \ar[dl] \ar[dr] \\
	D^\mu \ar[rr] && D_\mu}}
	\nonumber \\
	\qquad = -\frac{\partial}{\partial A^l}
	\xymatrix@C=1pc{(A^k \ar@<-0.5ex>[r] \ar@<0.5ex>[r] &
	A^l)}
	-\frac{\partial}{\partial \phi^l}
	\xymatrix@C=1pc{(A^k \ar@<-0.5ex>[r] \ar@<0.5ex>[r] &
	\phi^l)}
	-2 \vcenter{
	\xymatrix@C=0pc@R=0.8pc{& A^k \ar[dl] \ar[dr] \\
	\phi^l \ar[rr] && A^l}},
	\label{harddiagramident}
	\end{eqnarray}
where to prove the first step we expand the first term after the
equals sign by the chain rule and use the expansion of the vanishing
diagram $\xymatrix@C=1pc{(D^\mu \ar[r] & D_\mu) \ar@<-0.5ex>[r]
\ar@<0.5ex>[r] & A^k}$, and where the second step follows by using
(\ref{Dmudef}), expanding Poisson brackets in the $D^\mu$ coordinates,
and using the symmetries of the triangle diagram (see
(\ref{trianglediagram})) and (\Icontractionsign).  The first term on
the right hand side of (\ref{harddiagramident}) is of the type
indicated in (\ref{innocuousdiagram}) (hence, independent of origin of
the angles after averaging) and the second term vanishes upon
averaging.  Unfortunately, the third term still depends on the
origin of the angles, except in one dimension (see 
subsection~\ref{reduction1d}).

\subsection{Changing the independent variables}
\label{changingvariables}

By introducing a symplectic connection (Bayen \etal (1978), Fedosov
(1994)) the derivatives with respect to $z^\mu$ that appear in the
definition of a diagram (see \IMoyalnotation) can be replaced by
covariant derivatives that can be evaluated in any other coordinates,
of which the action-angle variables $D^\mu$ are an obvious choice.
Note that in action-angle coordinates, the operator $\partial/\partial
\phi^j$ is invariant under the change of origin
(\ref{angleoriginshift}), while the operator $\partial/\partial A^j$
depends on the origin (unless it is pulled out of an angle average, as
in (\ref{innocuousdiagram})).  Here we restrict attention to 
coordinates $D^\mu$, ignoring other possibilities, including various
non-coordinate frames.  The connection coefficients are required to
vanish in coordinates $z^\mu$ so that the original definitions of
diagrams remain unaltered (thus, the connection is flat), but they
take on non-zero values in the coordinates $D^\mu$.

The standard (nontensorial) transformation law for the connection
coefficients gives their values in coordinates $D^\mu$,
	\begin{equation}
	\Gamma^\mu{}_{\alpha\beta} =
	\frac{\partial D^\mu}{\partial z^\nu}
	\frac{\partial^2 z^\nu}{\partial D^\alpha 
	\partial D^\beta}=
	\frac{\partial z_\nu}{\partial D^\sigma}
	J^{\mu\sigma} 
	\frac{\partial^2 z^\nu}{\partial D^\alpha 
	\partial D^\beta},
	\label{Gammadef1}
	\end{equation}
where in the second equality we have used
	\begin{equation}
	\frac{\partial z^\mu}{\partial D^\alpha}
	J_{\mu\nu}
	\frac{\partial z^\nu}{\partial D^\beta}
	=J_{\alpha\beta}
	\label{Lagrangebracketxfm}
	\end{equation}
to express one Jacobian in terms of its inverse.  Only derivatives
with respect to $D^\mu$ appear in the final expression
(\ref{Gammadef1}).   The result is equivalent to
	\begin{equation}
	\Gamma_{\mu\alpha\beta} = 
	\frac{\partial z_\nu}{\partial D^\mu}
	\frac{\partial^2 z^\nu}{\partial D^\alpha 
	\partial D^\beta},
	\label{Gammadef2}
	\end{equation}
which is completely symmetric in indices $(\mu\alpha\beta)$, as may be
seen by differentiating (\ref{Lagrangebracketxfm}).

With these connection coefficients we can compute any diagram in terms
of covariant derivatives in action-angle coordinates.  For example, we
find
	\begin{equation}
	\xymatrix@C=1pc{D^\alpha \ar[r] & D^\beta \ar[r] &
	D^\gamma} =
	D^\alpha_{;\mu}\, J^{\mu\nu}\, D^\beta_{;\nu\sigma}
	\,J^{\sigma\tau} \,D^\gamma_{;\tau} =
	\Gamma^{\alpha\beta\gamma},
	\label{Gammadiagram}
	\end{equation}
which as in (\ref{3linearsymm}) shows the complete symmetry of this
diagram in indices $(\alpha\beta\gamma)$.  Other diagrams can be
expressed in terms of the connection coefficients, for example, by
inserting resolutions of the identity we find
	\begin{eqnarray}
	\xymatrix@C=1pc{D^\alpha \ar[r] & D^\beta \ar[r] &
	D^\mu \ar[r] & D^\nu} = 
	\Gamma^{\alpha\beta\sigma} \,
	\Gamma_\sigma{}^{\mu\nu}, \\
	\xymatrix@C=1pc{D^\alpha \ar@<-0.5ex>[r]
	\ar@<0.5ex>[r] & D^\beta} = 
	-\Gamma_\nu{}^{\alpha\mu} \,
	\Gamma_\mu{}^{\beta\nu}, \\
	\vcenter{
	\xymatrix@C=0pc@R=0.8pc{& D^\gamma \ar[dl] \\
	D^\alpha \ar[rr] && D^\beta \ar[ul]}} =
	\Gamma_\mu{}^{\alpha\nu} \,
	\Gamma_\nu{}^{\beta\sigma} \,
	\Gamma_\sigma{}^{\gamma\mu},
	\end{eqnarray}
etc.  As for the troublesome angle-dependent diagram in
(\ref{theresult}), it can be written in the interesting form,
	\begin{equation}
	\xymatrix@C=1pc{D^\mu \ar[r] & A^j \ar@<-0.5ex>[r] 
	\ar@<0.5ex>[r] & D_\mu} =
	\Gamma^{\mu\nu\sigma} \,
	\frac{\partial \Gamma_{\mu\nu\sigma}}
	{\partial \phi^j},
	\label{interestingform}
	\end{equation}
where we use (\ref{wheelidentity}).  Note that $\Gamma^{\mu\nu\sigma}
\, \Gamma_{\mu\nu\sigma} = 0$.  The derivatives of the connection
coefficients are not independent, due to the vanishing of the
curvature tensor.

\subsection{Reduction to the one-dimensional case}
\label{reduction1d}

It is not obvious that the quantization condition (\ref{theresult})
reduces to the one-dimensional formula (\IBSsecondorder) in the case
$N=1$.  To show that it does, we transform the multidimensional
formula and then specialize to $N=1$.  We begin with an identity,
	\begin{equation}
	\frac{\partial}{\partial A^l}
	\xymatrix@C=1pc{(A^j \ar[r] & A^k \ar[r] & A^l)} =
	-\xymatrix@C=1pc{A^j \ar@<-0.5ex>[r] \ar@<0.5 ex>[r] & A^k}
	-\frac{\partial}{\partial \phi^l}
	\xymatrix@C=1pc{(A^j \ar[r] & A^k \ar[r] & \phi^l)},
	\label{2A3Aidentity}
	\end{equation}
where the second term on the right vanishes upon $\phi$-averaging.  We
prove this by writing the left-hand side as $\xymatrix@C=1pc{\phi^l
\ar[r] & (A^j \ar[r] & A^l \ar[r] & A^k)}$, expanding this by the
chain rule (instead of using (\ref{wheelidentity})) to get three
diagrams, of which two can be combined into the first term on the
right hand side of (\ref{2A3Aidentity}), transformed by a resolution
of the identity.  The third diagram of the expansion can be
transformed by (\ref{wheelidentity}) into the second term on the right
hand side of (\ref{2A3Aidentity}).  Now using (\ref{harddiagramident})
and (\ref{2A3Aidentity}) we can transform the order $\hbar^2$ term of
(\ref{theresult}) (omitting the term $H^j_2$ and the $1/48$) into the form
	\begin{eqnarray}
	-\frac{\partial}{\partial A^l}
	\Biggl< \omega_{jk} \frac{\partial}{\partial A^m}
	\xymatrix@C=1pc{(A^k \ar[r] & A^l \ar[r] & A^m)}
	+ 2 \omega_{jkm}
	\xymatrix@C=1pc{(A^k \ar[r] & A^l \ar[r] & A^m)}
	\Biggr>_\phi
	\nonumber \\
	-2\omega_{jk} \Biggl<
	\vcenter{
	\xymatrix@C=0pc@R=0.8pc{& \phi^l \ar[dl] \ar[dr] \\
	A^k \ar[rr] && A^l}}
	\Biggr>_\phi.
	\label{xfm2ndorderterm}
	\end{eqnarray}
The final diagram of this expression contains the dependence on the
origin of the angles (before averaging).  

We now set $N=1$, dropping all Roman indices $jkl$, etc.  Then the
final diagram in (\ref{xfm2ndorderterm}) contains two identical ($A$)
entries, and therefore vanishes.  Then using the one-dimensional
version of (\ref{2A3Aidentity}), we can write the remainder of
(\ref{xfm2ndorderterm}) as
	\begin{equation}
	\frac{d}{dA}
	\left< \omega 
	\xymatrix@C=1pc{(A \ar@<-0.5ex>[r] \ar@<0.5ex>[r] & A)}
	-2\omega' 
	\xymatrix@C=1pc{(A \ar[r] & A \ar[r] & A)}\right>_\phi
	=\frac{d}{dA}
	\frac{1}{\omega}
	\left< 
	\xymatrix@C=1pc{H_0 \ar@<-0.5ex>[r] \ar@<0.5ex>[r] & H_0}
	\right>_\phi,
	\end{equation}
which agrees with the quantization rule cited in I if we expand,
$H=H_0 + \hbar^2 H_2 + \ldots$.   

\section{Conclusions}
\label{conclusions}

The calculations of I and this paper achieved a normal form in a
neighborhood of a stable fixed point of the Hamiltonian.  Other
authors (Helffer and Robert 1984, Colin de Verdi\`ere 2004) have
worked with an annular strip (in one-dimensional problems), that is, a
range of energies bounded away from the ground state.  This has the
advantage that one does not need to worry about what happens inside
the annular region: it may contain separatrices, nongeneric fixed
points, etc., and the Bohr-Sommerfeld rules are still valid.  In fact,
a WKB-Maslov kind of approach suggests that the asymptotic series for
the eigenvalues can be computed in the immediate neighborhood of a
single (quantized) torus.  In this sense, our approach has been more
global than it needs to be (for finding the eigenvalues).  On the
other hand, if one is studying normal forms, then a region containing
a fixed point is certainly one of the cases to be studied.  It is also
a necessary case, if one wishes to divide phase space into regions,
each of which can be transformed into some normal form.

The normal form transformations we have considered (in one and many
dimensions) are not the simplest.  That would be the case of a
contractible region of phase space in which the flow vector does not
vanish, that is, one not containing a critical (fixed) point of the
Hamiltonian.  In this case the quantum normal form is the same as the
classical, that is, the transformed Hamiltonian is just one of the
momenta, say, $p_1$.  Integrability (a set of other operators that
commute with $\hat H$) is not required.

As is well known, the torus quantization rule is not meaningful for
nonintegrable systems, and neither is the normal form transformation
of this paper.  That is, in a typical (nonintegrable) system, in a
neighborhood of a stable fixed point of the Hamiltonian, there is no
classical normal form, and hence no quantum normal form either, except
in the sense of approximations valid near the fixed point.  On the
other hand, the corresponding time-dependent problem, in which one
uses an extended phase space with $-H$ and $t$ as conjugate
coordinates, can be transformed into a simple normal form as described
in the preceding paragraph.

Conversely, instead of extending the phase space one can reduce it, in
cases of symmetry.  The case of $SO(3)$ symmetry is especially
important in practice (Littlejohn and Reinsch 1997).   It would be
interesting to study quantum normal forms on the reduced space, in
which the wave function is a section of a bundle, and their relation
to normal forms on the original phase space.

\appendix

\section{Notation and terminology}
\setcounter{section}{1}
\label{notation}

Wave functions in this paper are functions $\psi: {\mathbb R}^N \to
{\mathbb C}$, that is, they represent a scalar particle in
$N$-dimensional space.  The wavefunctions belong to the Hilbert space
$L^2({\mathbb R}^N)$, and operators such as ${\hat H}^j$ act on this
space.  Operators are indicated with hats, and their Weyl transforms
(or Weyl ``symbols'') are indicated by the same letter without the
hat, for example, ${\hat H}^j$ is the operator and $H^j$ its Weyl
symbol.

Latin indices $i$, $j$, etc.\ run from 1 to $N$.  The superscript or
subscript position of such indices, for example, $A^j$, $\omega_{ij}$,
has no significance.  Greek indices $\mu$, $\nu$, etc.\ run from 1 to
$2N$, for example, the coordinates on phase space are $z^\mu = (x_1,
\ldots, x_N, p_1, \ldots, p_N)$.  Superscript and subscript positions 
of Greek indices are meaningful, and indices are raised and lowered
with the Poisson tensor $J^{\mu\nu}$ and its inverse $J_{\mu\nu}$.
These are given in matrix form by
	\begin{equation}
	J^{\mu\nu} = \left(
	\begin{array}{cc}
	0 & I \\
	-I & 0 
	\end{array}
	\right),
	\qquad
	J_{\mu\nu} = \left(
	\begin{array}{cc}
	0 & -I \\
	I & 0
	\end{array}
	\right),
	\label{Jmunudef}
	\end{equation}
where $I$ is the $N \times N$ identity matrix.  Note the identity
(\Icontractionsign). 

As in I, we use a nonstandard notation for mappings between sets.  If
$A$ and $B$ are sets, we write $f:A \to B$ to indicate that the domain
of $f$ is some suitably chosen subset of $A$ (the standard notation
makes $A$ itself the domain).  See \Ifunctionnotation\ for notation
concerning functions and their derivatives.

In this paper, we denote the group of $N \times N$ integer matrices
with integer inverse by $GL(N,{\mathbb Z})$.  The determinant of such
a matrix is $\pm 1$.  If the determinant is restricted to $+1$, we
have the group $SL(N,{\mathbb Z})$.

\section{Symbols and functions of harmonic oscillators}
\label{hosymbols}

In this appendix we outline the proof in the multidimensional case
that an operator is a function of the set of harmonic oscillator
operators $\{ {\hat I}^j \}$ if and only if its symbol is a function
of the symbols $\{ I^j \}$ of these harmonic oscillators.  

The harmonic oscillators $\{ {\hat I}^j \}$ commute and possess a
simultaneous, nondegenerate eigenbasis, the usual basis $\{ \vert n_1
\ldots n_N \rangle \}$ of harmonic oscillator states.  Thus an
operator (say, $\hat A$) commutes with the operators $\{ {\hat I}^j
\}$ if and only if it is diagonal in this basis, which it does if and
only if it is a function of the operators $\{ {\hat I}^j \}$.  And
$\hat A$ commutes with the $\{ {\hat I}^j \}$ if and only if it
commutes with the unitary operators ${\hat U}^j(t) = \exp(-it
{\hat I}^j/\hbar)$,
	\begin{equation}
	{\hat U}^j(t) {\hat A}\, {\hat U}^j(t)^\dagger = {\hat A},
	\label{AUjcommutator}
	\end{equation}
for all $t$.  But according to Eq.~(\metaplecticcovariance) of
Littlejohn (1986) (expressing the well known metaplectic covariance of
the Weyl symbol, see Voros 1977, de Gosson 1997), when
(\ref{AUjcommutator}) is mapped into Weyl symbols we have
	\begin{equation}
	A\bigl(S^j(t)^{-1} z\bigr) = A(z),
	\label{AScommutator}
	\end{equation}
where $S(t)$ is the classical linear symplectic map generated by $I^j$
(that is, its Hamiltonian flow).  But this map is equivalent to
$\theta^j \mapsto \theta^j + t$, showing that the symbol $A(z)$ is
constant on each torus of the set of classical harmonic oscillators
$\{ I^j \}$.  This implies that the symbol $A$, regarded as a function
of $(\theta,I)$, is, in fact, independent of $\theta$, $A=A(I)$.
Conversely, if $A$ is such a function, then it satisfies
(\ref{AScommutator}) and hence the operator $\hat A$ satisfies
(\ref{AUjcommutator}) and therefore commutes with the ${\hat I}^j$.

\section{Normal forms for commuting, quadratic Hamiltonians}
\label{quadraticnormalform}

The proof of the theorem (\ref{matrixnormalform}) is a straightforward
but somewhat lengthy exercise in linear algebra, so here we shall
simply outline the highlights.  The proof relies critically on the
positive definiteness of the matrix $Q^1$ (without this condition the
conclusions do not hold).  The notation of this appendix is
independent of that of the rest of the paper, and will be explained as
we proceed.

Matrices $Q^j$ and $\Delta^j$ are as defined in
Sec.~\ref{thecalculation}.  In particular, the set $\{ Q^j \}$ of
real, symmetric, $2N \times 2N$ matrices is linearly independent, the
matrices commute in the sense of (\ref{QjQkcommutator}) or
(\ref{altQjQkcommutator}), and $Q^1$ is positive definite.  Matrix $J$
is the matrix with components $J^{\mu\nu}$, defined in \ref{Jmunudef},
and in this appendix we write $\omega=J^{-1}$ for its inverse (the
symplectic form).  Matrices $Q^j$ and $\omega$ specify bilinear forms
acting on ${\mathbb R}^{2N}$ or ${\mathbb C}^{2N}$.  We write
	\begin{equation}
	\eqalign{
	\omega(x,y) &= {\bar x}^T \omega y, \cr
	Q^j(x,y) &= {\bar x}^T Q^j y,}
	\label{bilinearforms}
	\end{equation}
for $x,y \in {\mathbb R}^{2N}$ or ${\mathbb C}^{2N}$, where ${\bar x}$
is the complex conjugate of $x$, so that the symbols $\omega$ and
$Q^j$ are used both for the matrices and for the corresponding forms.
Matrices $JQ^j$, on the other hand, specify maps $:{\mathbb R}^{2N}
\to {\mathbb R}^{2N}$ or $:{\mathbb C}^{2N} \to {\mathbb C}^{2N}$.

The matrix $JQ^1$ has purely imaginary eigenvalues that come in pairs
$\pm i\lambda$, where $\lambda >0$.  Let $(\lambda_1, \ldots,
\lambda_r)$ be the set of distinct positive numbers that specify
these eigenvalues.  Let $E_k \subset {\mathbb C}^{2N}$ be the
eigenspace of $JQ^1$ corresponding to eigenvalue $i\lambda_k$,
	\begin{equation}
	E_k = \{ x \in {\mathbb C}^{2N} | JQ^1 x = i\lambda_k x \},
	\label{Ekdef}
	\end{equation}
and let $n_k = \dim E_k$.  Then ${\bar E}_k$ is the eigenspace of
$JQ^1$ corresponding to eigenvalue $-i\lambda_k$.  It is linearly
independent of $E_k$, $E_k \cap {\bar E}_k = \{0\}$, and $\dim E_k =
\dim {\bar E}_k = n_k$.  Let $F^c_k = E_k \oplus {\bar E}_k \subset
{\mathbb C}^{2N}$, where the $c$ means ``complex,'' so that $\dim
F^c_k = 2n_k$.  Subspace $F^c_k$ possesses a real basis, and so
specifies a corresponding real subspace $F_k \subset {\mathbb R}^{2N}$
(the span of the real basis with real coefficients), with $\dim F_k =
2n_k$.  The subspaces $F_k$ are symplectic vector spaces under the
symplectic form $\omega$, and thus possess a basis $\{q_{k1}, \ldots,
q_{kn_k}, p_{k1}, \ldots, p_{kn_k}\}$ such that
	\begin{equation}
	\eqalign{
	\omega(q_{kl}, q_{km}) &= \omega(p_{kl}, p_{km}) = 0, \cr
	\omega(q_{kl}, p_{km}) &= -\omega(p_{kl}, q_{km}) 
	= - \delta_{lm},}
	\label{omegaab}
	\end{equation}
for $l,m = 1, \ldots, n_k$ and $k=1, \ldots, r$ (Arnold 1989, p.~219).
Moreover,
	\begin{equation}	
	F_1 \oplus \ldots \oplus F_r = {\mathbb R}^{2N},
	\label{Fdecomp}
	\end{equation}
so that
	\begin{equation}
	\sum_{k=1}^r n_k = N. 
	\label{sumnk}
	\end{equation}
This means that the matrix $JQ^1$ can be diagonalized by a linear
transformation, that is, its Jordan normal form has no
off-diagonal elements.   

The $\{q,p\}$ bases on the subspaces $F_k$ can be chosen so that the
matrix elements of $Q^1$ are a multiple of the identity on each
subspace,
	\begin{equation}
	\eqalign{
	Q^1(q_{kl}, q_{km}) &= Q^1(p_{kl}, p_{km}) = 
	\lambda_k\, \delta_{lm}, \cr
	Q^1(q_{kl}, p_{km}) &= Q^1(p_{kl}, q_{km})=0,}
	\label{Q1ab}
	\end{equation}
and so that the matrix elements of the other $Q^j$ are diagonal in the
form,
	\begin{equation}
	\eqalign{
	Q^j(q_{kl}, q_{km}) &= Q^j(p_{kl}, p_{km}) =
	\amatrix^j_{kl} \, \delta_{lm}, \cr
	Q^j(q_{kl}, p_{km}) &= Q^j(p_{kl}, q_{km})= 0,}
	\label{Qjab}
	\end{equation}
for $j>1$ (no sum on $l$), where the $\amatrix^j_{kl}$ are real numbers.

We create a single index for the $q$ and $p$ basis vectors by writing
	\begin{equation}
	\eqalign{
	q_i = (q_1, \ldots, q_N) &= (q_{11}, \ldots, q_{1n_1}, \ldots
	q_{r1}, \ldots, q_{rn_r}), \cr
	p_i =(p_1, \ldots, p_N) &= (p_{11}, \ldots, p_{1n_1}, \ldots
	p_{r1}, \ldots, p_{rn_r}),}
	\label{abrelabel}
	\end{equation}
for $i=1, \ldots, N$, and similarly we write
	\begin{equation}
	\lambda_i = (\lambda_1, \ldots, \lambda_1, \ldots,
	\lambda_r, \ldots, \lambda_r),
	\label{lambdarelabel}
	\end{equation}
with each $\lambda_k$ repeated $n_k$ times, and
	\begin{equation}
	\amatrix_{ji} = (\amatrix^j_{11}, \ldots, \amatrix^j_{1n_1}, \ldots,
	\amatrix^j_{r1}, \ldots, \amatrix^j_{rn_r}),
	\label{murelabel}
	\end{equation}
for $j>1$.  We also write $\amatrix_{1i} = \lambda_i$.  We arrange the
basis vectors columnwise in the order $(q,p)$ to create a $2N \times
2N$ matrix $S$.  Then (\ref{omegaab}) is equivalent to $S^T \omega S =
\omega$, so that $S$ is symplectic, and (\ref{Q1ab}) and (\ref{Qjab})
are equivalent to (\ref{matrixnormalform}), which was to be proven.
The $N \times N$ matrix $\amatrix_{ij}$ is nonsingular because the matrices
$Q^j$ are linearly independent.

\section{An example of the classical normal form transformation}
\label{normalformexample}

In this appendix we consider central force motion in two dimensions,
in which the Hamiltonian is
	\begin{equation}
	H({\bf x}, {\bf p}) =\frac{p^2}{2m} + V(r),
	\label{Hcentralforce}
	\end{equation}
where ${\bf x} = (x,y)$, ${\bf p} = (p_x,p_y)$, $r=|{\bf x}|$ and
$p=|{\bf p}|$.  We omit the hats, but $H$ can be either the quantum
Hamiltonian or its Weyl symbol (the classical Hamiltonian).  The
``other Hamiltonian'' is the $z$-component of the angular
momentum,
	\begin{equation}
	L({\bf x}, {\bf p}) = xp_y - yp_x.
	\label{Ldef}
	\end{equation}
The Weyl symbols of these two operators have no $\hbar$ dependence, so
the principal symbol and the whole symbol are identical, and we shall
simply write $H_j$ (with a subscript) instead of $H^j_0$.  Thus, $H_1
= H$ and $H_2 = L$.   

We assume the potential is smooth and has a minimum at $r=0$, so that
${\bf x} = {\bf p}=0$ is a stable fixed point of the classical motion.
We assume $V(0)=0$ so that $H$ and $L$ vanish at the fixed point.  We
write $\omega_0 = \sqrt{V''(0)/m}$ for the frequency of small
vibrations near the origin.  We transform to variables with balanced
units of $\hbox{\rm action}^{1/2}$ by writing ${\bf x} = {\bf
x}'/\sqrt{m\omega_0}$, ${\bf p} = \sqrt{m\omega_0} \, {\bf p}'$,
substituting these into the expressions for $H$ and $L$, and then
dropping the primes.  The function $L$ does not change under this
transformation, but $H$ becomes
	\begin{equation}
	H= \frac{\omega_0}{2} 
	[ p^2 + 2 U(r)],
	\label{Hbalanced}
	\end{equation}
where $U(r) = (1/\omega_0)V(r/\sqrt{m\omega_0})$.  Note that
$U''(0)=1$.  We transform (\ref{Hbalanced}) to polar coordinates
$(r,\theta)$, to obtain
	\begin{equation}
	H= \frac{\omega_0}{2}
	\left[ p_r^2 + \frac{L^2}{r^2} + 2U(r) \right]
	\label{Hpolar}
	\end{equation}
where $p_r = {\bf x} \cdot {\bf p}/r$ and $L=p_\theta$.  For positive
energies $E$ below the first separatrix (if there is one), the
classical motion in the $x$-$y$ plane takes place in an annular region
between two turning points $r_0$, $r_1$.

This region is the projection of the 2-torus in the 4-dimensional
phase space onto the $x$-$y$ plane.  The torus is characterized by the
values of the energy $E$ and angular momentum $L$ (in the latter case
we confuse a function with a value of a function).  The turning points
are the roots of (\ref{Hpolar}) obtained by setting $p_r=0$, that is,
the roots of
	\begin{equation}
	2E/\omega_0 = \frac{L^2}{r^2} + 2U(r).\
	\label{tpeqn}
	\end{equation}
There are two branches to the projection of the torus, identified by
the sign of $p_r$.  The branches are glued together at the radial
turning points, where $p_r=0$.  The angular momentum $L$ can take on
positive or negative values.  The inner turning point $r_0$ goes to
zero as $L$ passes through zero for fixed $E$, so the projection of
the torus closes up in the center and then opens up again.  Exactly at
$L=0$, the projection of the torus onto the $x$-$y$ plane is singular
at the origin (the inverse image of the origin is a circle on the
torus in phase space).

To study the normal form transformation for this system in a small
neighborhood of the fixed point, we expand (\ref{Hbalanced}) for small
$r$ to obtain,
	\begin{equation}
	H=\frac{\omega_0}{2} (p^2 + r^2) = 
	\frac{\omega_0}{2}
	(x^2 + y^2 + p_x^2 + p_y^2).
	\label{Hquadexp}
	\end{equation}
From this and (\ref{Ldef}) the matrices $Q^1$ and $Q^2$ defined by
(\ref{Qjdef}) are easily read off.  In particular, $Q^1 = \omega_0 I$
($I$ is the $4 \times 4$ identity matrix), so $Q^1$ is positive
definite.

The symplectic matrix $S$ in (\ref{matrixnormalform}) can be found by
various means, for example by noting that in this problem quadratic
Hamiltonians $H$ and $L$ belong to the Lie algebra of the $U(2)$
subgroup of $Sp(4)$, and using the theory of $SU(2)$ rotations to
transform $L$ into a diagonal form.  The $U(N)$ subgroup of $Sp(2N)$
is basic to the theory of the isotropic $N$-dimensional harmonic
oscillator (Littlejohn 1986, Cushman and Bates 1997).  We omit the
details and just quote the result, which we write as a linear
canonical transformation,
	\begin{equation}
	\left( \begin{array}{c}
	x \\ y \\ p_x \\ p_y
	\end{array} \right) = \frac{1}{\sqrt{2}}
	\left( \begin{array}{cccc}
	1 & 0 & 0 & -1 \\
	0 & 1 & -1 & 0 \\
	0 & 1 & 1 & 0 \\
	1 & 0 & 0 & 1
	\end{array} \right)
	\left( \begin{array}{c}
	x_1 \\ x_2 \\ p_1 \\ p_2
	\end{array}\right),
	\label{hotransformation}
	\end{equation}
where the matrix shown is the matrix $S$ of
(\ref{matrixnormalform}).  When (\ref{Hquadexp}) and (\ref{Ldef})
are transformed to the new coordinates, we find
	\begin{equation}
	\eqalign{
	H \circ Z^{-1} &= 
	\frac{\omega}{2}(x_1^2 + p_1^2 + x_2^2 + p_2^2) =
	\omega_0 (I_1 + I_2), \\
	L \circ Z^{-1} &=
	\frac{1}{2}(x_1^2 + p_1^2 - x_2^2 -p_2^2) = 
	I_1 - I_2,}
	\label{HLIxfm}
	\end{equation}
where $I_1$, $I_2$ are defined by (\ref{Ijdef}).  From this we can
read off the matrix $\amatrix_{jk}$ of (\ref{linearnormalform}), 
	\begin{equation}
	\amatrix_{jk} = \left( \begin{array}{cc}
	\omega_0 & \omega_0 \\
	1 & -1
	\end{array}\right).
	\label{mjkho}
	\end{equation}
	
The actions $A_j$ depend on the choice of basis contours on the tori.
The most obvious choice is $(\Gamma_r,\Gamma_\theta)$, the
contours whose projections onto the $x$-$y$ plane take place at
constant $\theta$ and $r$, respectively.  These are the contours that
emerge from the standard method of separating the Hamilton-Jacobi
equation in polar coordinates.  Contour $\Gamma_r$ goes the ``short
way'' around the torus, that is, from the inner to the outer radial
turning point and back again, with $p_r$ positive on the way out and
negative on the return.  Contour $\Gamma_\theta$ projects onto a
circle in the $x$-$y$ plane, traversed in a counterclockwise
direction, with fixed value of $(r,p_r)$.  The action associated with
$\Gamma_\theta$ is the angular momentum itself,
	\begin{equation}
	A_\theta = \frac{1}{2\pi} \oint p_\theta \, d\theta = L,
	\label{Athetadef}
	\end{equation}
while the radial action is
	\begin{equation}
	A_r = \frac{1}{2\pi} \oint p_r \, dr
	= \frac{1}{\pi} \int_{r_0}^{r_1}
	\sqrt{ \frac{2E}{\omega_0} - \frac{L^2}{r^2}
	- 2U(r)} \, dr.
	\label{Ardef}
	\end{equation}

A tricky aspect about these contours and actions is that the contour
$\Gamma_r$ is not continuous as $L$ passes through zero for fixed $E$,
that is, $\Gamma_r$ for small positive $L$ is not close (in phase
space) to the contour $\Gamma_r$ for small negative $L$.  Instead, the
contour at small negative $L$ that is close to $\Gamma_r$ for small
positive $L$ is $\Gamma_r - \Gamma_\theta$, that is, the contour,
continuously tracked as $L$ passes through zero from positive values,
picks up an extra (backwards) loop around the origin.  As a result,
$(\Gamma_r,\Gamma_\theta)$ and $(A_r,A_\theta)$ do not constitute a
smooth assignment of basis contours and actions in the region of
interest, but rather only over the two subregions $L>0$ and $L<0$
separately.

In the quadratic approximation (\ref{Hquadexp}) to the Hamiltonian (an
isotropic harmonic oscillator) the radial action $A_r$ can be
evaluated explicitly.  In this case the energy and angular momentum
obey the bounds $-E/\omega_0 \le L \le +E/\omega_0$, and the radial
turning points are
	\begin{equation}
	r_{0,1} = \sqrt{ E/\omega_0 \pm
	\sqrt{ (E/\omega_0)^2 - L^2 }}.
	\label{turningpoints}
	\end{equation}
Then direct integration of (\ref{Ardef}) gives
	\begin{equation}
	A_r = \frac{1}{2} \left( \frac{E}{\omega_0} - |L| \right).
	\label{Arho}
	\end{equation}
The absolute value signs in the second term is an indication that
$\Gamma_r$ and hence $A_r$ are not continuous at $L=0$.

For the quadratic approximations, a different contour on the tori is
conceptually useful, namely, the one generated by $H$ itself.  This is
possible because the isotropic harmonic oscillator possesses only
periodic orbits.  Call this contour $\Gamma_H$.  By direct examination
of the elliptical orbit in the $x$-$y$ plane, it is obvious that
	\begin{equation}
	\Gamma_H = 2 \Gamma_r \pm \Gamma_\theta,
	\label{GammaHeqn}
	\end{equation}
where the $\pm$ is the sign of $L$.   Thus,
	\begin{equation}
	A_H = \frac{1}{2\pi} \oint_{\Gamma_H} 
	{\bf p} \cdot d{\bf x} = 2 A_r + |L|.
	\label{AHeqn1}
	\end{equation}
The contour $\Gamma_H$ is continuous as $L$ passes through zero; one
function of the absolute value sign in (\ref{AHeqn1}) is to cancel the
discontinuity in the term $2A_r$.  The action $A_H$ is easily
evaluated in rectangular coordinates with time as the parameter of the
loop by using Hamilton's equations of motion.  The result is
	\begin{equation}
	A_H = \frac{1}{2\pi} \int_0^{2\pi/\omega_0}
	{\bf p} \cdot \dot{\bf x} \, dt = E/\omega_0.
	\label{AHeqn}
	\end{equation}
This integral is much easier than the one that led to (\ref{Arho}),
but combined with (\ref{AHeqn1}) it gives the same result for $A_r$.

The orbits of the Hamiltonian $H$ are closed curves only in the
quadratic approximation (\ref{Hquadexp}), not for the full Hamiltonian
(\ref{Hbalanced}).  Nevertheless, the closed contour $\Gamma_H$, defined on
tori near the fixed point, can be extended continously to closed
contours on tori at larger values of $E$, where (\ref{GammaHeqn}) and
hence (\ref{AHeqn1}) are still valid.

Let us choose the basis contours to be $(\Gamma_1,\Gamma_2) =
(\Gamma_r, \Gamma_\theta)$, so that $(A_1,A_2) = (A_r, L)$.  Then
contours $(C_1, C_2)$, the basis contours on the tori of harmonic
oscillators $(I_1, I_2)$, when mapped by $Z$ back onto contours on the
tori of $(H,L)$, can be worked out in the $(x,y,p_x,p_y)$ coordinates
by using (\ref{hotransformation}).  For example, to get $C_1$, we may
write $x_1 = \sqrt{2I_1} \sin \theta_1$, $p_1 =
\sqrt{2I_1} \cos \theta_1$, $x_2 = 0$, $p_2 = \sqrt{2I_2}$, substitute
into (\ref{hotransformation}), and let $\theta_1$ range from 0 to
$2\pi$.  By plotting these contours in the $x$-$y$ plane and keeping
track of the sign of $p_r$, we obtain the following relations among
contours:
	\begin{equation}
	\left.\eqalign{
	C_1 &= \Gamma_r + \Gamma_\theta, \\
	C_2 &= \Gamma_r,}\right\}
	\quad L>0; \qquad
	\left.\eqalign{
	C_1 &= \Gamma_r, \\
	C_2 &= \Gamma_r - \Gamma_\theta,}\right\}
	\quad L<0,
	\label{hocontourmaps}
	\end{equation}
where $\Gamma_r$ and $\Gamma_\theta$ really mean $Z \circ \Gamma_r$
and $Z \circ \Gamma_\theta$.  Integrating the symplectic form over
these contours, we obtain,
	\begin{equation}
	\left.\eqalign{
	I_1 &= A_r + L, \\
	I_2 &= A_r,}\right\}
	\quad L>0; \qquad
	\left.\eqalign{
	I_1 &= A_r, \\
	I_2 &= A_r -L,}\right\}
	\quad L<0,
	\label{hoactionmaps}
	\end{equation}
where $I_1$ and $I_2$ really mean $I_1 \circ Z$ and $I_2 \circ Z$.
These relations may also be obtained (more easily) by combining
(\ref{Arho}) and (\ref{HLIxfm}).  From them it is easy to obtain the
matrix $\nu_{jk}$ of (\ref{Zcontourmap}).  We find,
	\begin{equation}
	\nu = \left(
	\begin{array}{cc}
	0 & 1 \\
	1 & -1
	\end{array} \right),
	\quad L>0; \qquad 
	\nu=\left(
	\begin{array}{cc}
	1 & 0 \\
	1 & -1
	\end{array}\right),\quad
	L<0.
	\label{honumatrices}
	\end{equation}

\section{Existence of the generators $G_{2n}$}
\label{Gnexistence}

The notation of this appendix differs slightly from that of the main
text, in that we write $\hat M^j$ for some intermediate step in the
second stage transformations,
	\begin{equation}
	{\hat M}^j = {\hat U}_{2n-2} \ldots {\hat U}_4 {\hat U}_2
                   {\hat K}^j {\hat U}_2^\dagger
		{\hat U}_4^\dagger \ldots
	       {\hat U}_{2n-2}^\dagger,
	\label{intermedMdef}
	\end{equation}
where $n \ge 1$ (rather than the final product as $n \to \infty$, as
in the main text).  We also write
	\begin{equation}
	{\hat M}^{\prime j} =
	{\hat U}_{2n} \, {\hat M}^j \, {\hat U}_{2n}^\dagger
	\label{Mprimedef}
	\end{equation}
for the next stage in the transformation.  Symbols $M^j$ and
$M^{\prime j}$ of ${\hat M}^j$ and ${\hat M}^{\prime j}$ are expanded
in $\hbar$ as in (\ref{Mjexpansion}).  We assume that $M^j_0 = K^j_0$
and that $M^j_{2m}$, expressed as a function of $\theta$ and $I$, is
independent of $\theta$ for $0 \le m \le n-1$.  Under these
assumptions we will show that $G_{2n}$ exists such that $M^{\prime
j}_{2m} = M^j_{2m}$ for $0 \le m \le n-1$ and such that $M^{\prime
j}_{2n}$ is independent of $\theta$ (that is, to two higher orders in
$\hbar$ than the previous step).  Since the stated assumptions hold
for $n=1$, by induction they hold for any $n \ge 1$.

The unitary operator ${\hat U}_{2n}$ is given in terms of its
generator ${\hat G}_{2n}$ by (\IUndef), where here we assume that
$G_{2n}$ is independent of $\hbar$.  Thus, (\ref{Mprimedef}) has the
expansion,
	\begin{equation}
	{\hat M}^{\prime j} = {\hat M}^j -i\hbar^{2n-1}
	[{\hat G}_{2n}, {\hat M}^j ] - \frac{1}{2}
	\hbar^{4n-2} [ {\hat G}_{2n}, [ {\hat G}_{2n}, 
	{\hat M}^j ]] + \ldots.
	\label{Mprimejexpn}
	\end{equation}
Upon transcribing to symbols and expanding as in (\ref{Mjexpansion})
we find $M^{\prime j}_{2m} = M^j_{2m}$ for $0 \le m \le n-1$, and
	\begin{equation}
	M^{\prime j}_{2n} = M^j_{2n} + \{ G_{2n} , M^j_0 \}
	= M^j_{2n} + \omega_{jk} 
	\frac{\partial G_{2n}}{\partial \theta^k}.
	\label{Mprimej2neqn}
	\end{equation}
Taking the averaged part of this we obtain $M^{\prime j}_{2n} = {\bar
M}^j_{2n}$, that is, $M^{\prime j}_{2n}$ is independent of $\theta$.
Taking the oscillatory part we obtain a differential equation for
$G_{2n}$,
	\begin{equation}
	\frac{\partial G_{2n}}{\partial \theta^k} =
	-(\omega^{-1})_{jk} \, {\tilde M}^j_{2n}.
	\label{G2neqn}
	\end{equation}
A solution for this exists if the second derivatives
	\begin{equation}
	\frac{\partial^2 G_{2n}}{\partial \theta^k \partial \theta^l}
	=-(\omega^{-1})_{kj} \, 
	\frac{\partial M^j_{2n}}{\partial \theta^l}
	\label{G2n2ndderivs}
	\end{equation}	
are symmetric in $k$ and $l$, where we drop the tilde on $M^j_{2n}$
since the $\theta^l$ derivative of the averaged part vanishes.
Rearranging, we see that a solution $G_{2n}$ of (\ref{G2neqn}) exists
if and only if
	\begin{equation}
	-\omega_{lj} \, 
	\frac{\partial M^k_{2n}}{\partial \theta^j}
	+\omega_{kj} \,
	\frac{\partial M^l_{2n}}{\partial \theta^j} =0,
	\end{equation}
that is, if
	\begin{equation}
	\{ M^k_0, M^l_{2n} \} + \{ M^k_{2n}, M^j_0 \} =0.
	\label{G2nic}
	\end{equation}

However, since the operators ${\hat M}^j$ are unitarily equivalent to
the ${\hat H}^j$ and since the latter commute, we have $[ {\hat M}^k,
{\hat M}^l ]=0$.  Transcribing this to symbols and using
(\IMoyalcommutator), we obtain the hierarchy of equations,
	\begin{eqnarray}
	\{ M^k_0, M^l_0 \} =0, 
	\nonumber \\
	\{ M^k_0, M^l_2 \} + \{ M^k_2, M^l_0 \} 
	-\frac{1}{24} \{ M^k_0, M^l_0 \}_3 =0, 
	\nonumber \\
	\{ M^k_0, M^l_4 \} + \{ M^k_2, M^l_2 \} +
	\{ M^k_4, M^l_0 \} -\frac{1}{24}
	\{ M^k_0, M^l_2 \}_3 \nonumber \\
	\qquad - \frac{1}{24}
	\{ M^k_2, M^l_0 \} + \frac{1}{1920}
	\{ M^k_0, M^l_0 \}_5 =0,
	\label{Mcommhier}
	\end{eqnarray}
etc.
	
Now let $A$ and $B$ be any two functions of $I$ that are independent
of $\hbar$.  Then all the odd order Moyal brackets $\{ A,B \}_{2n+1}$
vanish.  This follows from the fact that the corresponding operators
$\hat A$ and $\hat B$ are functions of $\hat I$ (see \ref{hosymbols}),
and hence commute, $[ {\hat A}, {\hat B} ]=0$.  Expanding the symbol
version of this commutator as in (\IMoyalcommutator) then yields $\{
A, B \}_{2n+1} =0$ for all $n \ge 0$.

Applying this theorem to (\ref{Mcommhier}) and using the fact that
$M^j_{2m}$ is independent of $\theta$ for $0 \le m \le n-1$, we see
that all terms in the first $n-1$ equations of the hierarchy vanish,
while the $n$-th one gives precisely the integrability condition
(\ref{G2nic}).  Thus, $G_{2n}$ exists.

\section{An antiderivative for (\ref{indefintegral})}
\label{antiderivative}

We wish to evaluate the integral
	\begin{equation}
	\int_0^1 d\epsilon \, 
	\xymatrix@C=1pc{(G_\epsilon \ar@<1.0ex>[r]
	\ar[r] \ar@<-1.0ex>[r] & A^j_\epsilon)}
	\circ Z^{-1}_\epsilon,
	\label{hardintegral}
	\end{equation}
to within terms that vanish on taking a $\theta$-average (not a
$\phi$-average, as in (\ref{fj2def}), because we have removed the
$\circ Z$ seen in (\ref{indefintegral})).  We will write the integrand
in the form $(d/d\epsilon)(F_\epsilon \circ Z^{-1}_\epsilon)$ for some
function $F_\epsilon$ to be determined.  According to (\IDDepsilondef)
and (\IaltDDepsilondef), $F_\epsilon$ satisfies,
	\begin{equation}
	\frac{D F_\epsilon}{D\epsilon} = 
	\frac{d F_\epsilon}{d\epsilon} - \{ F_\epsilon, G_\epsilon \}
	= \xymatrix@C=1pc{G_\epsilon \ar@<1.0ex>[r]
	\ar[r] \ar@<-1.0ex>[r] & A^j_\epsilon}.
	\label{antiderivdef}
	\end{equation}
This is the sense in which we shall regard $F_\epsilon$ as the
antiderivative of the integrand of (\ref{hardintegral}).  

For notational convenience we now drop the $\epsilon$-subscripts on
$A^j_\epsilon$, $\phi^j_\epsilon$, $D^\mu_\epsilon$, $Z_\epsilon$,
$F_\epsilon$ and $G_\epsilon$ until further notice.  That is, $A^j$,
$\phi^j$, $D^\mu$, $Z$, etc., now stand for quantities evaluated at
any $\epsilon$ (not just $\epsilon=1$, as in the main text or in
Table~\ref{table}).  

The functions $F$ we are interested in are made up of diagrams
involving the action-angle variables $D^\mu$.  The derivative
$dF/d\epsilon$ is evaluated by the chain rule, in which
	\begin{eqnarray}
	\frac{d D^\mu}{d\epsilon} &= 
	\frac{d}{d\epsilon}(E^\mu \circ Z) =
	(E^\mu_{,\nu} \circ Z) 
	\frac{dZ^\nu}{d\epsilon}
	=(E^\mu_\nu \circ Z) \{ Z^\nu, G \} 
	= \{ E^\mu \circ Z, G \} 
	\nonumber \\
	&= \{ D^\mu, G\}
	=\xymatrix@C=1pc{D^\mu \ar[r] & G},
	\label{Dmuderiv}
	\end{eqnarray}
where we use (\ref{Aphiepsilondef}) in the first equality, the chain
rule in the second, (\IZsubepsilonode) in the third, and the chain
rule again in the fourth.  See the first three terms of
(\ref{DDepsilonexpansion}) for an example of the application of this
formula.

Thus, taking the derivative $D/D\epsilon$ of a diagram creates a
series of new diagrams, each with one extra arrow and one extra copy
of $G$ compared to the original diagram.  Since the integrand of
(\ref{hardintegral}) has one $G$, one copy of $A^j$ and three arrows,
the suggestion is that the antiderivative must be a diagram with $A^j$
and two arrows.  Unfortunately, this cannot be correct, since any
diagram with an arrow that begins and terminates on the same operand
($A^j$ in this case) vanishes, by the antisymmetry of $J^{\mu\nu}$.
The basic logic is correct, however, it is just that we must take into
account the possibility of introducing a resolution of the identity
(see \ref{resolution}) into the original integrand, which adds one copy
each of $D^\mu$ and $D_\mu$, and one extra arrow.  Thus, the original
integrand can be regarded as a diagram with $D^\mu$, $D_\mu$, $A^j$,
$G$, and four arrows.  Therefore candidates for the antiderivative
would involve diagrams with $D^\mu$, $D_\mu$, $A^j$ and three arrows
(and no $G$).  It would also be possible to introduce more than one
resolution of the identity, but it turns out that one suffices for our
purposes.

Therefore candidate diagrams for the antiderivative include the
following:
	\begin{eqnarray}
	\xymatrix@C=1pc{D^\mu \ar[r] & 
	A^j \ar@<0.5ex>[r] \ar@<-0.5ex>[r]
	& D_\mu}, \qquad
	\xymatrix@C=1pc{A^j \ar[r] & 
	D^\mu \ar@<0.5ex>[r] \ar@<-0.5ex>[r]
	& D_\mu},
	\nonumber \\
	\xymatrix@C=1pc{A^j \ar@<0.5ex>[r] \ar@<-0.5ex>[r]
	& D^\mu \ar[r] & D_\mu}, \qquad
	\vcenter{
	\xymatrix@C=0pc@R=0.8pc{& A^j \\
	D^\mu \ar[ur] \ar[rr] && D_\mu \ar[ul]}},
	\label{candidatediagrams}
	\end{eqnarray}	
where we do not count as distinct any diagrams that can be obtained
from the ones given by reversing the direction of arrows or by using
(\Icontractionsign).  Let us call these diagrams $T_1$, $T_2$, $T_3$
and $T_4$, in the order given.

These diagrams are not all independent.  First we note that the
Poisson bracket $\{ D^\mu, D^\nu \} = J^{\mu\nu}$ is a constant, so
any derivatives of it vanish.  The action $A^j$ can be regarded
as a component of some $2N$-vector $D^\nu$, so the same applies to
Poisson brackets $\{ D^\mu, A^j \}$, etc.  

These constant Poisson brackets lead to various identities, such as
	\begin{eqnarray}
	0 &= 
	\xymatrix@C=1pc{A^j \ar@<0.5ex>[r] \ar@<-0.5ex>[r]
	& (D^\mu \ar[r] & D_\mu)} =
	\xymatrix@C=1pc{A^j \ar@<0.5ex>[r] \ar@<-0.5ex>[r]
	& D^\mu \ar[r] & D_\mu} +
	2 \vcenter{
	\xymatrix@C=0pc@R=0.8pc{& A^j \ar[dl] \ar[dr] \\
	D^\mu \ar[rr] && D_\mu }} 
	\nonumber \\
	&+\xymatrix@C=1pc{D^\mu \ar[r] & D_\mu 
	& A^j\ar@<0.5ex>[l] \ar@<-0.5ex>[l] }
	=2(T_3+T_4),
	\label{PBident1}
	\end{eqnarray}
and
	\begin{eqnarray}
	0 &=
	\xymatrix@C=1pc{D_\mu \ar@<0.5ex>[r] \ar@<-0.5ex>[r]
	& (A^j \ar[r] & D^\mu)} =
	\xymatrix@C=1pc{D_\mu \ar@<0.5ex>[r] \ar@<-0.5ex>[r]
	& A^j \ar[r] & D^\mu} +
	2 \vcenter{
	\xymatrix@C=0pc@R=0.8pc{& D_\mu \ar[dl] \ar[dr]  \\
	A^j \ar[rr] && D^\mu }} 
	\nonumber \\
	&+\xymatrix@C=1pc{A^j \ar[r] & D^\mu & D^\mu 
	\ar@<0.5ex>[l] \ar@<-0.5ex>[l]} =
	-T_1 + 2T_4 +T_2,
	\label{PBident2}
	\end{eqnarray}
where we expand using the chain rule for diagrams.  We see that only
two of the diagrams (\ref{candidatediagrams}) are independent, say,
$T_1$ and $T_2$.

We take the first of these $T_1$ as a guess for the desired
antiderivative, and compute,
	\begin{eqnarray}
	\frac{D}{D\epsilon} 
	&\xymatrix@C=1pc{(D^\mu \ar[r] & 
	A^j \ar@<0.5ex>[r] \ar@<-0.5ex>[r]
	& D_\mu)} =
	\xymatrix@C=1pc{(D^\mu \ar[r] & G) \ar[r] & A^j 
	\ar@<-0.5ex>[r] \ar@<0.5ex>[r] & D_\mu} +
	\xymatrix@C=1pc{D^\mu \ar[r] & (A^j \ar[r]
	&G) \ar@<-0.5ex>[r] \ar@<0.5ex>[r] & D_\mu}
	\nonumber \\
	&+\xymatrix@C=1pc{D^\mu \ar[r] & A^j
	\ar@<-0.5ex>[r] \ar@<0.5ex>[r] &(D^\mu \ar[r]
	&G)} -
	\xymatrix@C=1pc{(D^\mu \ar[r] & A^j
	\ar@<-0.5ex>[r] \ar@<0.5ex>[r] & D_\mu) \ar[r]
	&G},
	\label{DDepsilonexpansion}
	\end{eqnarray}
using (\ref{antiderivdef}) and (\ref{Dmuderiv}).  The four diagrams on
the right hand side can be expanded by the chain rule, resulting in
fourteen diagrams altogether, but ten of these cancel just by changing
the directions of arrows or using (\Icontractionsign).  The remaining
four diagrams add up to a continuation of the previous equation,
	\begin{eqnarray}
	= 
	\xymatrix@C=1pc{D^\mu \ar[r] & A^j \ar[r] & G
	\ar@<-0.5ex>[r] \ar@<0.5ex>[r] & D_\mu} 
	-2\raise2.7pc\vbox{
	\xymatrix@C=1pc@R=1pc{
	&D_\mu \\
	D^\mu \ar[r]
	& G \ar[u] \ar [r]
	& A^j \ar[ul]}}
	-\raise2.7pc\vbox{
	\xymatrix@C=1pc@R=1pc{
	&D_\mu \\
	D^\mu \ar[r] 
	& G \ar@<-0.5ex>[u] \ar@<0.5ex>[u] \ar[r]
	& A^j}}
	\nonumber \\
	\qquad -\xymatrix@C=1pc{
	D^\mu \ar[r] & A^j \ar@<-0.5ex>[r] \ar@<0.5ex>[r]
	& G \ar[r] & D_\mu}.
	\label{DDepsilonexpansion1}
	\end{eqnarray}

We notice the following facts about this result.  First, the final
diagram is the one we want, $\xymatrix@C=1pc{G \ar@<-1ex>[r] \ar[r]
\ar@<1ex>[r] & A^j}$, slightly disguised with a resolution of the
identity.  Next, we are allowed to split off an exact
$\phi$-derivative (i.e., a $\phi_\epsilon$-derivative, which turns
into a $\theta$-derivative when composed with $Z_\epsilon^{-1}$).
But if $X$ is any diagram, then
	\begin{equation}
	\xymatrix@C=1pc{A^j \ar[r] & (X)} 
	=-\frac{\partial X}{\partial \phi^j},
	\label{Aarrowphideriv}
	\end{equation}
as we see by expanding the Poisson bracket in $(\phi,A)$ coordinates.
So to create an exact $\phi^j$-derivative, we must create diagrams of
the form $\xymatrix@C=1pc{A^j \ar[r] & (X)}$.  Third, to create such
diagrams we can employ identities like (\ref{PBident1}) and
(\ref{PBident2}), which follow by using some constant Poisson bracket
as a part of a larger diagram.  In the present case, useful identities
follow from the vanishing expressions,
	\begin{equation}
	\xymatrix@C=1pc{(D^\mu \ar[r] & A^j) \ar[r] & G
	\ar@<-0.5ex>[r] \ar@<0.5ex>[r] & D_\mu},
	\qquad
	\xymatrix@C=1pc{D^\mu \ar[r] & G
	\ar@<-0.5ex>[r] \ar@<0.5ex>[r] & (D_\mu \ar[r] &
	A^j)}.
	\label{vanishingdiagrams}
	\end{equation}
By using these we obtain finally,
	\begin{equation}
	\frac{D}{D\epsilon}
	\xymatrix@C=1pc{(D^\mu \ar[r] & 
	A^j \ar@<0.5ex>[r] \ar@<-0.5ex>[r]
	& D_\mu)} =
	2 \raise0.35pc\vbox{
	\xymatrix@C=1pc{G \ar@<-1ex>[r] \ar[r]
	\ar@<1ex>[r] & A^j}}
	-\frac{\partial}{\partial \phi^j} 
	\xymatrix@C=1pc{(D^\mu \ar[r] & G
	\ar@<-0.5ex>[r] \ar@<0.5ex>[r] & D_\mu)},
	\label{theantideriv}
	\end{equation}
showing that the first guess for the desired antiderivative worked
(the other one, the second term in (\ref{candidatediagrams}), called
$T_2$ above, does not yield the desired integrand). 

Now restoring the $\epsilon$'s and using (\ref{antiderivdef}), we have
	\begin{eqnarray}
	2\int_0^1 d\epsilon \, 
	\xymatrix@C=1pc{(G_\epsilon \ar@<1.0ex>[r]
	\ar[r] \ar@<-1.0ex>[r] & A^j_\epsilon)}
	\circ Z^{-1}_\epsilon = 
	\left. 
	\xymatrix@C=1pc{(D^\mu_\epsilon \ar[r] & A^j_\epsilon
	\ar@<-0.5ex>[r] \ar@<0.5ex>[r] & D_{\epsilon\mu})}
	\circ Z^{-1}_\epsilon \right|^1_0
	\nonumber \\
	\qquad +
	\int^1_0 d\epsilon \, \left[
	\frac{\partial}{\partial \phi^j_\epsilon}
	\xymatrix@C=1pc{(D^\mu_\epsilon \ar[r] & G_\epsilon
	\ar@<-0.5ex>[r] \ar@<0.5ex>[r] & D_{\epsilon\mu})}
	\right]\circ Z^{-1}_\epsilon
	\nonumber \\
	=\xymatrix@C=1pc{(D^\mu \ar[r] & A^j
	\ar@<-0.5ex>[r] \ar@<0.5ex>[r]  & D_\mu)}
	\circ Z^{-1} 
	-\xymatrix@C=1pc{E^\mu \ar[r] & I^j
	\ar@<-0.5ex>[r] \ar@<0.5ex>[r]  & E_\mu}
	\nonumber \\
	\qquad +\frac{\partial}{\partial \theta^j} \int^1_0 d\epsilon \,
	\xymatrix@C=1pc{(D^\mu_\epsilon \ar[r] & G_\epsilon
	\ar@<-0.5ex>[r] \ar@<0.5ex>[r] & D_{\epsilon\mu})}
	\circ Z^{-1}_\epsilon.
	\label{doingtheintegral}
	\end{eqnarray}
The middle term on the right hand side vanishes, since $I^j$ is a
quadratic polynomial in $z^\mu$ with three arrows attached to it, and
in the final integral we have used the composition with
$Z_\epsilon^{-1}$ to convert the $\phi^j_\epsilon$ derivative into a
$\theta^j$ derivative, which can be taken out of the integral.
Dividing the result by 2 and composing with $Z$ gives
(\ref{indefintegral}).

\section{Resolution of the identity}
\label{resolution}

The technique we call ``inserting a resolution of the identity'' can
be used to break a bond of a diagram, resulting in a new diagram (or
product of diagrams) with one extra arrow and an extra copy of $D^\mu$
and $D_\mu$.  In this appendix, $A$, $B$ and $C$ are any functions,
while $D^\mu$ and $D_\mu$ are as in the main text (see (\ref{Dmudef})).

Consider a diagram (or product of diagrams) of the form
$\xymatrix@C=1pc{D_\mu \ar[r] & X \ar[r] & D^\mu}$, where $X$
stands for a single function, a diagram with more than one function
connected by arrows (where the arrows coming into or out of $X$ are
attached anywhere on the diagram), or a product of such
diagrams.  Examples are $\xymatrix@C=1pc{D_\mu \ar[r] & A \ar[r] &
D^\mu}$, $\xymatrix@C=1pc{D_\mu \ar[r] & A \ar[r] & B \ar[r] & D^\mu}$,
or $\xymatrix@C=1pc{(D_\mu \ar[r] & A)} \xymatrix@C=1pc{(B 
\ar@<-0.5ex>[r] \ar@<0.5ex>[r] & C \ar[r] & D^\mu)}$.  Then we have
	\begin{equation}
	\xymatrix@C=1pc{D_\mu \ar[r] & X \ar[r] & D^\mu} =
	\frac{\partial D_\mu}{\partial z^\beta} \,
	J^{\beta\sigma} \, {}_\sigma X_\tau \, 
	J^{\tau\alpha} \,
	\frac{\partial D^\mu}{\partial z^\alpha}.
	\label{roiparadigm}
	\end{equation}
But since the components of the symplectic form are preserved under a
canonical coordinate transformation, we have
	\begin{equation}
	\frac{\partial D^\mu}{\partial z^\alpha}
	\frac{\partial D_\mu}{\partial z^\beta} =
	\frac{\partial D^\mu}{\partial z^\alpha}
	\, J_{\mu\nu} \,
	\frac{\partial D^\nu}{\partial z^\beta} =
	J_{\alpha\beta},
	\label{Lagrangebracket}
	\end{equation}
and therefore,
	\begin{equation}
	\xymatrix@C=1pc{D_\mu \ar[r] & X \ar[r] & D^\mu} =
	J^{\tau\sigma} \, {}_\sigma X_\tau =
	\raise 2.25pc \vbox{
	\xymatrix@C=1pc@R=1pc{& & \\
	& X \ar `r[ur] `[u] `[l] `[] [] & }}.
	\label{roi}
	\end{equation}
For example,
	\begin{eqnarray}
	\xymatrix@C=1pc{D_\mu \ar[r] & A \ar[r] & D^\mu}
	=\raise 2.1pc \vbox{
	\xymatrix@C=1pc@R=1pc{& & \\
	& A \ar `r[ur] `[u] `[l] `[] [] & }} = 0, \\
	\xymatrix@C=1pc{D_\mu \ar[r] & A \ar[r] 
	& B \ar[r] & D^\mu} =
	\xymatrix@C=1pc{A \ar@<-0.5ex>[r] & B \ar@<-0.5ex>[l]} =
	-\xymatrix@C=1pc{A \ar@<-0.5ex>[r] \ar@<0.5ex>[r] & B}, \\
	\xymatrix@C=1pc{(D_\mu \ar[r] & A)} \xymatrix@C=1pc{(B 
	\ar@<-0.5ex>[r] \ar@<0.5ex>[r] & C \ar[r] & D^\mu)}=
	\xymatrix@C=1pc{B \ar@<-0.5ex>[r] \ar@<0.5ex>[r] & 
	C \ar[r] & A}.
	\label{roiegs}
	\end{eqnarray}
	
\section{Diagrammatic identities}
\label{identities}

In this appendix we present some diagrammatic identities involving the
action-angle variables $D^\mu$.  

For a linear diagram with three $D$'s, we have
	\begin{equation}
	\xymatrix@C=1pc{D^\mu \ar[r] & D^\nu \ar[r] & D^\alpha}=
	\hbox{\rm symmetric in $(\mu\nu\alpha)$}.
	\label{3linearsymm}
	\end{equation}
This follows by expanding vanishing diagrams such as
$\xymatrix@C=1pc{(D^\mu \ar[r] & D^\nu) \ar[r] & D^\alpha}$ by the
chain rule, and by inverting the directions of all arrows.

Similarly, the triangle diagram,
	\begin{equation} 
	\xymatrix@C=0pc@R=0.8pc{& D^\gamma \ar[dl] \\
	D^\alpha \ar[rr] && D^\beta \ar[ul]} 
	\label{trianglediagram}
	\end{equation}
can be shown to be symmetric (antisymmetric) under even (odd)
permutations of $(\alpha\beta\gamma)$. 

For a linear diagram with four $D$'s, let us abbreviate
	\begin{equation}
	\xymatrix@C=1pc{D^\mu \ar[r] & D^\nu \ar[r] & 
	D^\alpha \ar[r] & D^\beta}=[\mu\nu\alpha\beta].
	\label{4linearabbrev}
	\end{equation}
Then we have the symmetries,
	\begin{equation}
	[\mu\nu\alpha\beta] = [\nu\mu\alpha\beta] =
	[\mu\nu\beta\alpha] = -[\alpha\beta\mu\nu],
	\label{4linearsymm}
	\end{equation}
which follow by the same techniques used in proving (\ref{3linearsymm}).

Finally, we have the identity,
	\begin{equation}
	\xymatrix@C=1pc{D^\mu \ar[r] & (D^\alpha \ar[r] &
	D^\beta \ar[r] & D^\gamma)} = \vcenter{
	\xymatrix@C=1pc@R=0.3pc{ & & D^\beta \\
	D^\alpha & D^\mu \ar[l] \ar[ur] \ar[dr] \\
	& & D^\gamma}}.
	\label{wheelidentity}
	\end{equation}
We prove this by directly expanding the diagram on the left by the
chain rule and then using the expansion of the vanishing diagram
$\xymatrix@C=1pc{D^\alpha \ar[r] & (D^\mu \ar[r] & D^\beta) \ar[r] &
D^\gamma}$ and the identities (\ref{4linearsymm}).  Note that both
sides of (\ref{wheelidentity}) are manifestly symmetric in
$(\alpha\beta\gamma)$.

\section*{References}
\begin{harvard}

\item[] Abraham R and Marsden J E 1978 {\it Foundations of Mechanics}
(Reading, Massachusetts: Benjamin/Cummings)

\item[] Alber M S, Luther G G, Marsden J E, and Robbins J M 1998 {\it
Physica} {\bf 123D} 271

\item[] Arnold V I 1989 {\it Mathematical Methods of Classical
Mechanics} (New York: Springer-Verlag)

\item[] Bayen F, Flato M, Fronsdal C, Lichnerowicz A and Sternheimer
D 1978 {\it Ann. Phys.} {\bf 111} 61

\item[] Berry M V and Tabor M 1976 {\it Proc.\ Roy.\ Soc.\ Lond.\ A}
{\bf 349} 101

\item[] Brillouin M L 1926 {\it J.\ Phys.} {\bf 7} 353

\item[] Cargo Matthew, Gracia-Saz Alfonso, Littlejohn R G, Reinsch M W
and Rios P de M 2004 {\it J. Phys.\ A} {\bf 38} 1977

\item[] Charbonnel, A-M 1986 {\it Integral Equations Oper.\ Theory}
{\bf 9} 502

\item[] Child M S 1998 {\it J. Phys.\ A} {\bf 31} 657

\item[] Colin de Verdi\`ere Y 2004 Bohr-Sommerfeld rules to all orders 
{\it Preprint} http://www-fourier.ujf-grenoble.fr/~ycolver/ebk.ps

\item[] Cushman R H and Bates L 1997 {\it Global Aspects of Classical 
Integrable Systems} (Basel: Birkh\"auser Verlag)

\item[] Cushman R H and Duistermaat J J 1988 {\it Bull.\ Am.\ Math.\
Soc.} {\bf 19} 475

\item[] Cushman R H, Dullin H R, Giacobbe A, Holm D D, Joyeux M, Lynch
P, Sadovski\'\i\ D A, and Zhilinski\'\i\ B I 2004 {\it Phys.\ Rev.\
Lett.} {\bf 93} 024302-1

\item[] de Gosson M 1997 {\it Maslov Classes, Metaplectic
Representation and Lagrangian Quantization} (New York: Wiley)

\item[] Duistermaat J J 1980 {\it Commun.\ Pure Appl.\ Math.} {\bf 33}
687

\item[] Eckhardt B and Wintgen D {\it J. Phys. A} {\bf 24} 4335

\item[] Einstein A 1917 {\it Verh.\ dt.\ Phys.\ Ges.} {\bf 19} 82

\item[] Eliasson L H 1990 {\it Comment.\ Math.\ Helvetici} {\bf 65} 4

\item[] Fedosov B V 1994 {\it J. Differential Geometry} {\bf 40} 213

\item[] Fomenko A T 1991 {\it Advances in Soviet Mathematics} vol~6
(Providence, Rhode Island: American Mathematical Society)

\item[] Foxman J A and Robbins J M 2004 {\it preprint} math-ph/0411017

\item[] Gracia-Saz A 2005 {\it preprint} math.QA/0411163 

\item[] Helffer B and Robert D 1984 {\it Ann.\ Inst.\ H. Poincar\'e
Phys.\ Th\'eor.} {\bf 41} 291

\item[] Keller J B 1958 {\it Ann.\ Phys.} {\bf 4} 180

\item[] Kirillov A A 1976 {\it Elements of the Theory of
Representations} (New York:  Springer-Verlag)

\item[] Littlejohn R G 1986 {\it Phys. Reports} {\bf 138} 193

\item[] Littlejohn R G and Reinsch M 1997 {\it Rev.\ Mod.\ Phys.} {\bf
69} 213

\item[] Littlejohn R G and Robbins J M 1987 {\it Phys.\ Rev.\ A} {\bf
36} 2953

\item[] Marsden J E and Ratiu T S 1999 {\it Introduction to Mechanics 
and Symmetry} 2nd ed (New York: Springer-Verlag)

\item[] Maslov V P and Fedoriuk M V 1981 {\it Semi-Classical
Approximations in Quantum Mechanics} (Dordrecht: D. Reidel)

\item[] Percival I C 1973 {\it J.\ Phys.\ B} {\bf 6} L229

\item[] Voros A 1977 {\it Ann. Inst. Henri Poincar\'e} {\bf 4} 343

\item[] Weinstein A 1977 {\it Lectures on Symplectic Manifolds,
Regional Conference Series in Mathematics} {\bf 29} (Providence, Rhode
Island: American Mathematical Society).

\end{harvard}

\Figures

\begin{figure}
\caption{\label{EMmapsinfinitesimal} The two energy-momentum maps (the
$H_0$-map and the $I$-map) are related by map $Z$ and the matrix $a$.
The figure illustrates the case $N=2$.}
\end{figure}

\begin{figure}
\caption{\label{mapsroi} Maps associated with the region of interest,
$R$.  The four axes in the diagrams on the left represent ${\mathbb
R}^{2N}$, and the two axes in the diagrams on the right represent
${\mathbb R}^N$.}
\end{figure}

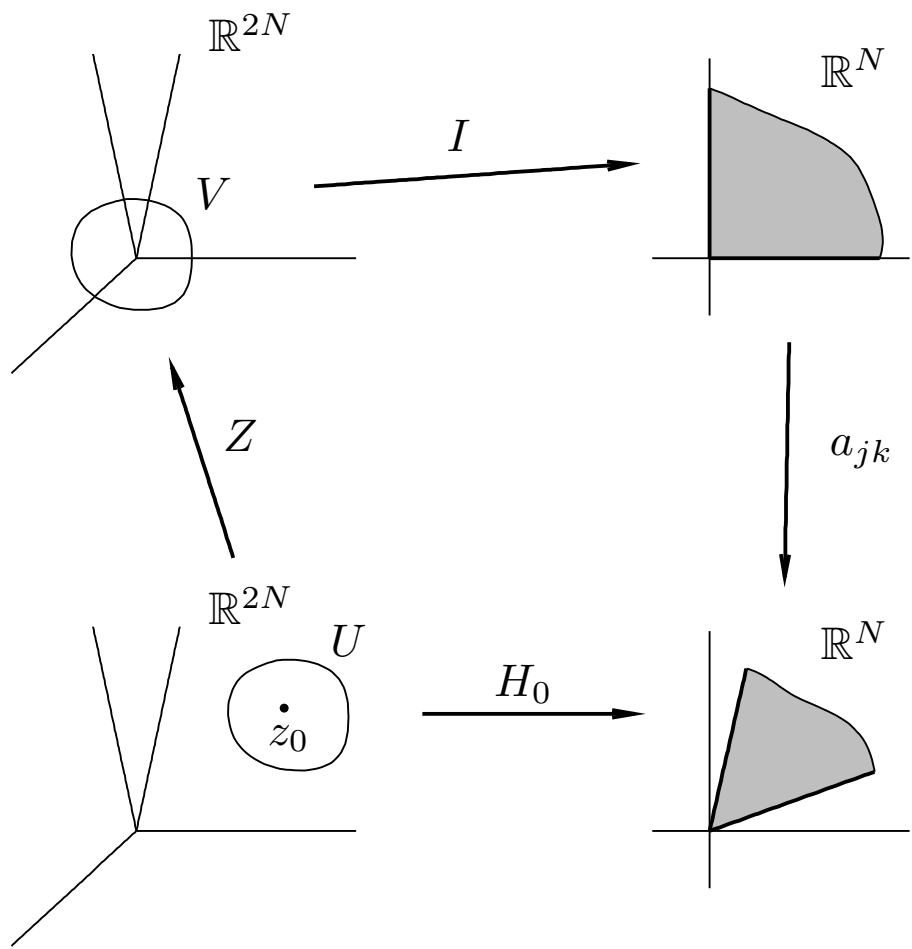
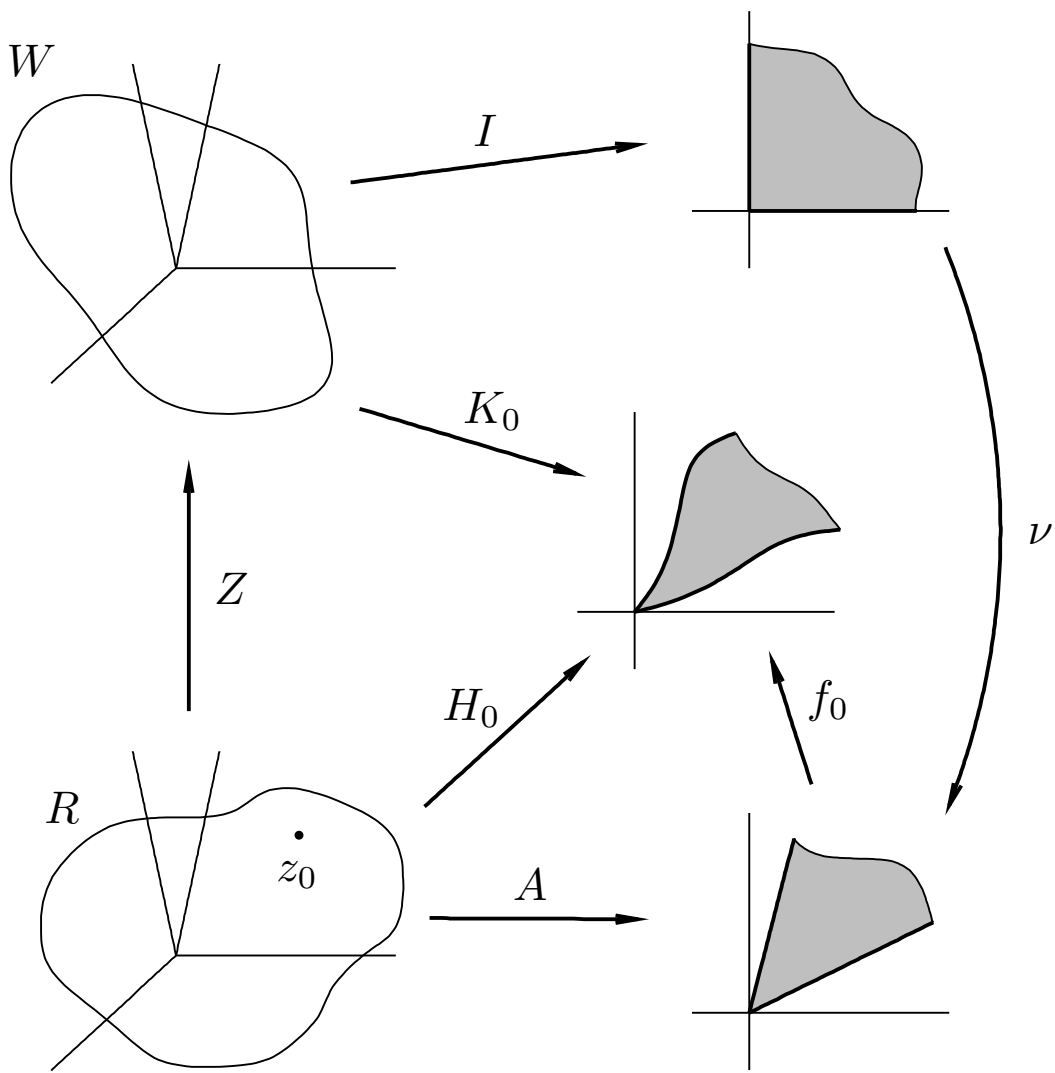
\end{document}